\documentclass[10pt,double]{IEEEtran}

\usepackage{amsmath}
\usepackage{amssymb}
\usepackage{amsthm}
\usepackage{latexsym}
\usepackage{verbatim}
\usepackage{graphicx}

\newtheorem{thm}{Theorem}

\newtheorem{problem}{Problem}

{\begin{list}               %    with flush left bullets.
    {$\bullet$ \hfill}{
        \setlength{\leftmargin}{\parindent}
        \setlength{\parsep}{0.04\baselineskip}
        \setlength{\itemsep}{0.5\parsep}
        \setlength{\labelwidth}{\leftmargin}
        \setlength{\labelsep}{0em}}
    }
{\end{list}}

\providecommand{\cref}[1]{Chapter~\ref{chap:#1}}

\providecommand{\R}{\ensuremath{\mathbb{R}}}

\renewcommand{\S}{\ensuremath{\mathbb{S}}}

\providecommand{\abs}[1]{\left|#1\right|}
\providecommand{\norm}[1]{\left\lVert#1\right\rVert}
\providecommand{\inprod}[1]{\left\langle#1\right\rangle}
\providecommand{\set}[1]{\left\{#1\right\}}

\providecommand{\bydef}{\overset{\text{def}}{=}}
\providecommand{\diag}{\mathop{\mathrm{diag}}}
\providecommand{\rank}{\mathop{\mathrm{rank}}}

\renewcommand{\vec}[1]{\ensuremath{\boldsymbol{#1}}}
\providecommand{\mat}[1]{\ensuremath{\boldsymbol{#1}}}
\providecommand{\wh}[1]{\ensuremath{\widehat{#1}}}
\providecommand{\wt}[1]{\ensuremath{\widetilde{#1}}}

\providecommand{\mA}{\mat{A}} \providecommand{\mB}{\mat{B}}
 \providecommand{\mD}{\mat{D}}
 \providecommand{\mH}{\mat{H}}
\providecommand{\mI}{\mat{I}} \providecommand{\mJ}{\mat{J}}

\providecommand{\mM}{\mat{M}}  
\providecommand{\mQ}{\mat{Q}} \providecommand{\mR}{\mat{R}}
\providecommand{\mS}{\mat{S}} \providecommand{\mU}{\mat{U}} 
\providecommand{\mV}{\mat{V}} \providecommand{\mT}{\mat{T}}
\providecommand{\mW}{\mat{W}}
\providecommand{\mSigma}{\mat{\Sigma}}
 \providecommand{\mG}{\mat{G}}

\providecommand{\mX}{\mat{X}}\providecommand{\mY}{\mat{Y}}
\providecommand{\mZ}{\mat{Z}}

\providecommand{\va}{\vec{a}} \providecommand{\vb}{\vec{b}}
 \providecommand{\vd}{\vec{d}}
\providecommand{\ve}{\vec{e}}

 \providecommand{\vn}{\vec{n}} 
 \providecommand{\vp}{\vec{p}}
 \providecommand{\vr}{\vec{r}}
\providecommand{\vs}{\vec{s}}
\providecommand{\vt}{\vec{t}}

\providecommand{\vx}{\vec{x}} \providecommand{\vy}{\vec{y}}
 
 \providecommand{\vzero}{\vec{0}}

\pdfpageattr {/Group << /S /Transparency /I true /CS /DeviceRGB>>} 

\usepackage{bbm}
\usepackage{bm}
\usepackage{dsfont}
\usepackage{algorithm}
\usepackage{subfigure}
\usepackage{multicol}
\usepackage{algpseudocode}
\usepackage{xcolor}
\usepackage[pagebackref=false,
            citecolor=blue,
            urlcolor=blue,
            colorlinks=true]{hyperref}
\usepackage{hyperref}
\usepackage[framemethod=tikz]{mdframed}
\usepackage{booktabs}
\usepackage{multirow} 
\usepackage{tabularx}
\usepackage{pifont}
\usepackage[numbered]{mcode}
\usepackage{mathtools}
\usepackage{float}
\DeclareRobustCommand{\mtx}{\texorpdfstring{$\mR = \left[\substack{\ \ 0 \ 1 \\ -1 \ 0}\right]$}{0 \ 1; -1 0}}

\newcommand{\EDM}{\ensuremath{\mathrm{edm}}}
\newcommand{\EDMset}{\ensuremath{\mathbb{EDM}}}
\newcommand{\EDMgram}{\ensuremath{\mathcal{K}}}
\newcommand{\GramSmall}{\ensuremath{\mathcal{G}}}
\newcommand{\vone}{\ensuremath{\vec{\mathit{1}}}}
\renewcommand{\vzero}{\ensuremath{\vec{\mathit{0}}}}
\newcommand{\mzero}{\ensuremath{\mat{\mathit{0}}}}
\newcommand{\T}{\ensuremath{\top}}
\newcommand{\trace}{\ensuremath{\mathop{\mathrm{trace}}}}
\newcommand{\mDelta}{\bm{\mathit{\Delta}}}
\newcommand{\mLambda}{\bm{\mathit{\Lambda}}}
\renewcommand{\diag}{\mathop{\mathrm{diag}}}

\newcommand{\affdim}{\ensuremath{\mathrm{affdim}}}
\newcommand{\SVD}{\ensuremath{\mathrm{SVD}}}
\newcommand{\EVD}{\ensuremath{\mathrm{EVD}}}
\DeclareMathOperator*{\argmin}{arg\,min}
\DeclareMathOperator*{\argmax}{arg\,max}

\definecolor{border_color}{rgb}{0.0, 0.0, 0.0}
\definecolor{fill_color}{rgb}{0.9, 0.9, 0.9}
\newmdenv[innerlinewidth=0.25pt,
          linecolor=border_color, 
	      backgroundcolor=fill_color,  
          innerleftmargin=6pt, 
          innerrightmargin=6pt,
          innertopmargin=6pt,
          innerbottommargin=6pt]{spmagbox}

\definecolor{nice_blue}{rgb}{0.5 0.1 0.9}
\newcommand{\rev}[1]{{#1}}

\makeatletter
\newlength{\continueindent}
\renewenvironment{algorithmic}[1][0]%
   {%
   \edef\ALG@numberfreq{#1}%
   \def\@currentlabel{\theALG@line}%
   \setcounter{ALG@line}{0}%
   \setcounter{ALG@rem}{0}%
   \let\\\algbreak%
   \expandafter\edef\csname ALG@currentblock@\theALG@nested\endcsname{0}%
   \expandafter\let\csname ALG@currentlifetime@\theALG@nested\endcsname\relax%
   \begin{list}%
      {\ALG@step}%
      {%
      \rightmargin\z@%
      \itemsep\z@ \itemindent\z@ \listparindent2em%
      \partopsep\z@ \parskip\z@ \parsep\z@%
      \labelsep 0.5em \topsep 0.2em%\skip 1.2em 
      \ifthenelse{\equal{#1}{0}}%
         {\labelwidth 0.5em}%
         {\labelwidth 1.2em}%
       \leftmargin\labelwidth \addtolength{\leftmargin}{\labelsep}
      \ALG@tlm\z@%
      }%
      \parshape 2 \leftmargin \linewidth \continueindent \dimexpr\linewidth-\continueindent\relax
   \setcounter{ALG@nested}{0}%
   \ALG@beginalgorithmic%
   }%
   {% end{algorithmic}
   \ALG@closeloops%
   \expandafter\ifnum\csname ALG@currentblock@\theALG@nested\endcsname=0\relax%
   \else%
      \PackageError{algorithmicx}{Some blocks are not closed!!!}{}%
   \fi%
   \ALG@endalgorithmic%
   \end{list}%
   }%
\makeatother
\title{{\fontsize{1.18cm}{1em}\selectfont Euclidean Distance Matrices} \\ \huge{Essential Theory, Algorithms and Applications}}
\author{Ivan Dokmani\'{c}, Reza Parhizkar, Juri Ranieri and Martin Vetterli
\thanks{Ivan Dokmani\'{c}, Juri Ranieri and Martin Vetterli are with the School of Computer and Communication
Sciences, Ecole Polytechnique F\'{e}d\'{e}rale de Lausanne (EPFL), CH-1015
Lausanne, Switzerland (e-mails: \{ivan.dokmanic, juri.ranieri@epfl.ch,
martin.vetterli\}@epfl.ch). Reza Parhizkar is with macx red AG (Financial services), CH-6300 Zug, Switzerland (e-mail: reza.parhizkar@gmail.com).}%
\thanks{Ivan Dokmani\'c and Juri Ranieri were supported by the ERC Advanced
Grant---Support for Frontier Research---SPARSAM, Nr: 247006. Ivan Dokmani\'c
was also supported by the Google PhD Fellowship.} }

\begin{document}
\maketitle

% \tableofcontents

% \linespread{1.45}

\begin{abstract}
Euclidean distance matrices (EDM) are matrices of squared distances between
points. The definition is deceivingly simple: thanks to their many useful
properties they have found applications in psychometrics, crystallography,
machine learning, wireless sensor networks, acoustics, and more. Despite the
usefulness of EDMs, they seem to be insufficiently known in the signal
processing community. Our goal is to rectify this mishap in a concise
tutorial.

We review the fundamental properties of EDMs, such as  rank or
(non)definiteness. We show how various EDM properties can be used to design
algorithms for completing and denoising distance data. Along the way, we
demonstrate applications to microphone position calibration, ultrasound
tomography, room reconstruction from echoes and phase retrieval. By spelling
out the essential algorithms, we hope to fast-track the readers in applying
EDMs to their own problems. Matlab code for all the described algorithms, and
to generate the figures in the paper, is available online. Finally, we suggest
directions for further research.
\end{abstract}

%============================================================================

\section{Introduction}

Imagine that you land at Geneva airport with the Swiss train schedule but no
map. Perhaps surprisingly, this may be sufficient to reconstruct a rough (or
not so rough) map of the Alpine country, even if the train times poorly
translate to distances or some of the times are unknown. The way to do it is
by using Euclidean distance matrices (EDM): for a quick illustration, take a
look at the ``\textbf{Swiss Trains}'' box.

An EDM is a matrix of squared Euclidean distances between points in a
set.\footnote{\rev{While there is no doubt that a Euclidean distance matrix
should contain Euclidean distances, and not the squares thereof, we adhere to
this semantically dubious convention for the sake of compatibility with most
of the EDM literature. Often, working with squares does simplify the
notation.}} We often work with distances because they are convenient to
measure or estimate. In wireless sensor networks for example, the sensor nodes
measure
\rev{received signal strengths} of the packets sent by other nodes, or time-%
of-arrival (TOA) of pulses emitted by their neighbors \cite{Patwari:2005kc}.
Both of these proxies allow for distance estimation between pairs of nodes,
thus we can attempt to reconstruct the network topology. This is often termed
\emph{self-localization}
\cite{Alfakih1999,dohetry2001,Biswas2004}. The molecular conformation problem
is another instance of a distance problem \cite{Havel1985281},
and so is reconstructing a room's geometry from echoes
\cite{Dokmanic:2013dz}. Less
obviously, sparse phase retrieval \cite{Ranieri:2013tx} can be converted to a
distance problem, and addressed using EDMs.

\begin{figure}
\centering
\includegraphics[width=3.5in]{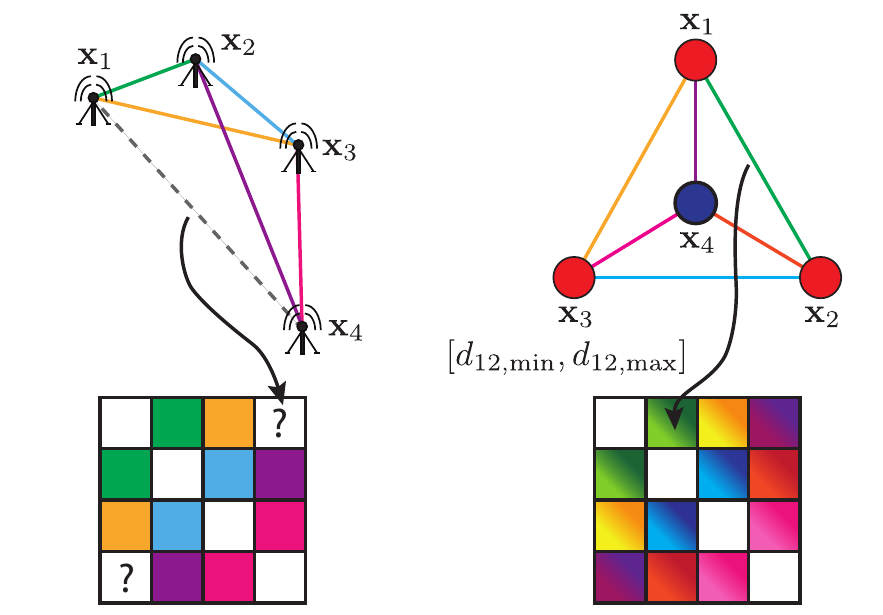}
\caption{Two real-world applications of EDMs. Sensor network localization
from estimated pairwise distances is illustrated on the left, with one
distance missing because the corresponding sensor nodes are too far apart to
communicate. In the molecular conformation problem on the right, we aim to
estimate the locations of the atoms in a molecule from their pairwise
distances. Here, due to the inherent measurement uncertainty, we know the
distances only up to an interval.}
  \vspace{-2mm}
\label{fig:intro}
\end{figure}

Sometimes the data are not metric, but we seek a
metric representation, as it happens commonly in psychometrics
\cite{torgerson1952}. As a matter of fact, the psychometrics community is
at the root of the development of a number of tools related to EDMs, including
multidimensional scaling (MDS)---the problem of finding the best point set
representation of a given set of distances. More abstractly, people are
concerned with EDMs for objects living in high-dimensional vector spaces, such
as images \cite{Weinberger2004}.

EDMs are a useful description of the point sets, and a starting point for
algorithm design. A typical task is to retrieve the original point
configuration: it may initially come as a surprise that this requires no more
than an eigenvalue decomposition (EVD) of a symmetric
matrix.\footnote{\rev{Because the EDMs are symmetric, we choose to use EVDs
instead of singular value decompositions. That the EVD is much more efficient for
symmetric matrices was suggested to us by one of the reviewers of the initial
manuscript, who in turn received the advice from the numerical analyst Michael
Saunders.}} In fact, the majority of Euclidean distance problems require the
reconstruction of the point set, but always with one or more of the following
twists:
\begin{enumerate}
	\item Distances are noisy,
	\item Some distances are missing,
	\item Distances are unlabeled.
\end{enumerate}
For two examples of applications requiring a solution of EDM problems with
different complications, see Fig. \ref{fig:intro}.
%----------------------------------------------------------------------------
% Swiss trains box
%----------------------------------------------------------------------------
\begin{figure}[t]
\fontfamily{lmss}\selectfont
\begin{spmagbox}
\textbf{Swiss Trains (Swiss Map Reconstruction)}
\\
\fontsize{9pt}{10pt}
\begin{center}
	\includegraphics[width=3.3in]{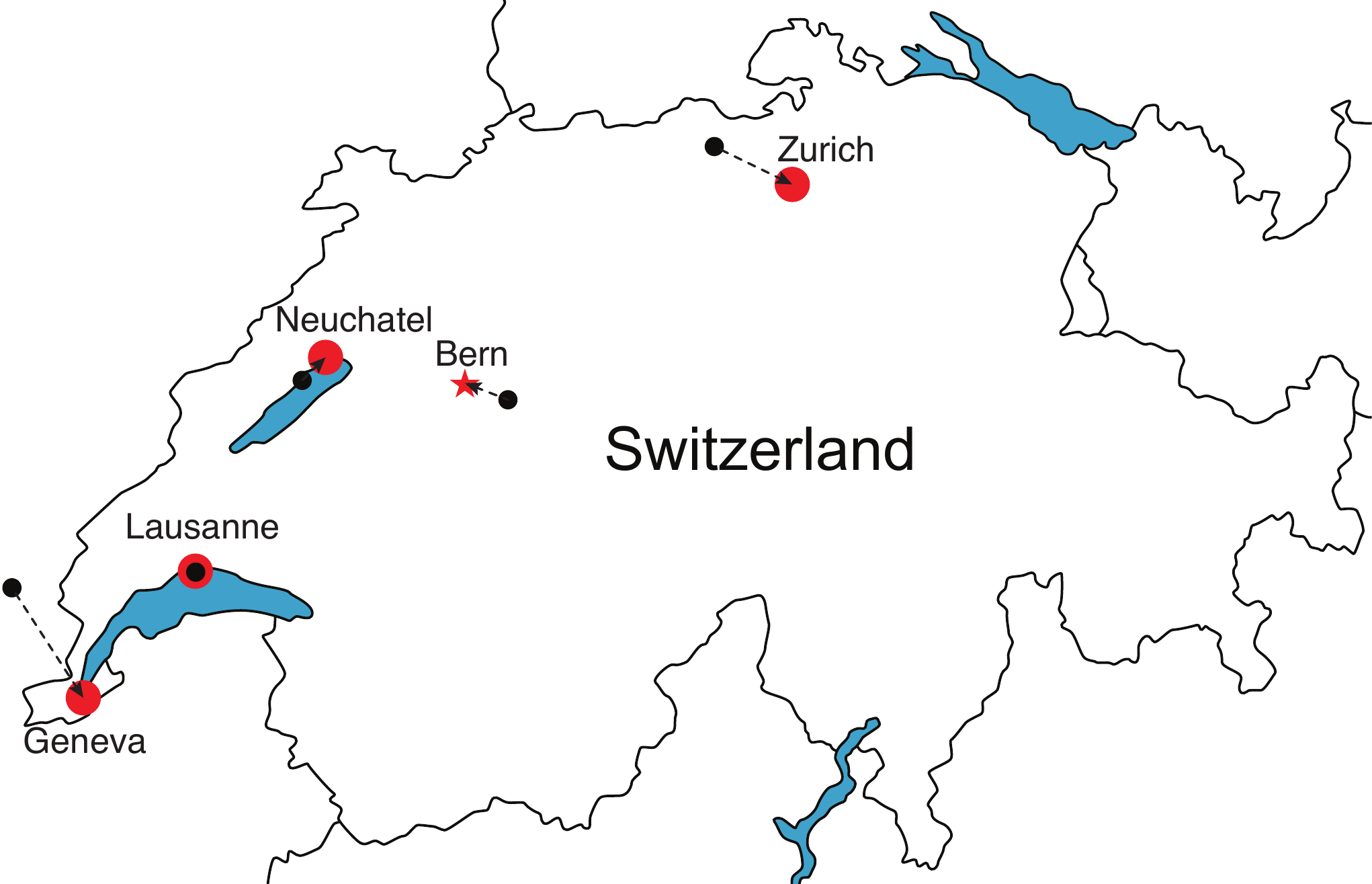}
\end{center}
\vspace{8mm}

\selectfont
Consider the following matrix of times in minutes it takes to travel by train
between some Swiss cities:
\begin{equation*}
\bordermatrix{     & \text{L} & \text{G} & \text{Z} & \text{N} & \text{B} \cr
\text{Lausanne}    & 0   & 33  & 128 & 40 & 66  \cr
\text{Geneva}      & 33  & 0   & 158 & 64 & 101 \cr
\text{Z\"urich}    & 128 & 158 & 0   & 88 & 56  \cr
\text{Neuch\^atel} & 40  & 64  & 88  & 0  & 34  \cr
\text{Bern}        & 66  & 101 & 56  & 34 & 0   \cr
}
\end{equation*}
The numbers were taken from the Swiss railways timetable. The matrix
was then processed using the classical MDS algorithm (Algorithm
\ref{alg:classical_mds}), which is basically an EVD. The obtained city
configuration was rotated and scaled to align with the actual map. Given all
the uncertainties involved, the fit is remarkably good. Not all trains drive
with the same speed; they have varying numbers of stops, railroads are not
straight lines (lakes and mountains). This result may be regarded as
anecdotal, but in a fun way it illustrates the power of the EDM toolbox.
Classical MDS could be considered the simplest of the available tools, yet it
yields usable results with erroneous data. On the other hand, it might be that
Swiss trains are just so good. 
\end{spmagbox}
\end{figure}
%----------------------------------------------------------------------------

There are two fundamental problems associated with distance geometry
\cite{Liberti:2012ut}: (i) given a matrix, determine whether it is an EDM,
(ii) given a possibly incomplete set of distances, determine whether there
exists a configuration of points in a given embedding dimension---dimension of the
smallest affine space comprising the points---that generates the distances.

% Sometimes the algorithms directly exploit the structure of the matrix,
% typically its rank. Sometimes the matrix is a useful abstraction, only
% indirectly exploited by the algorithms.

\subsection{Prior Art}

The study of point sets through pairwise distances, and so of EDMs, can be
traced back to the works of Menger
\cite{Menger:1928bc}, Schoenberg \cite{Schoenberg:1935dk},
Blumenthal \cite{Blumenthal:1953ie}, and Young and Householder
\cite{young1938}.

An important class of EDM tools was initially developed for the purpose of
data visualization. In 1952, Torgerson introduced the notion of MDS
\cite{torgerson1952}. He used distances to quantify the
\emph{dissimilarities} between pairs of objects that are not necessarilly
vectors in a metric space. Later in 1964, Kruskal suggested the notion of
\emph{stress} as a measure of goodness-of-fit for non-metric data
\cite{kruskal1964}, again representing experimental dissimilarities between
objects. 

A number of analytical results on EDMs were developed by Gower
\cite{gower1982, gower1}. In his 1985 paper \cite{gower1}, he gave a complete
characterization of the EDM rank. Optimization with EDMs requires good
geometric intuitions about matrix spaces. In 1990, Glunt \cite{Glunt1990} and
Hayden \cite{Hayden1990} with their co-authors provided insights into the
structure of the convex cone of EDMs. An extensive treatise on EDMs with many
original results and an elegant characterization of the EDM cone is given by
Dattorro \cite{Dattorro:2011wa}.

In early 1980s, Williamson, Havel and W\"{u}thrich developed the idea of
extracting the distances between pairs of hydrogen atoms in a protein, using
nuclear magnetic resonance (NMR). The extracted distances were then used to
reconstruct 3D shapes of molecules\footnote{W\"{u}thrich received the Nobel
Prize for chemistry in 2002.} \cite{Havel1985281}. The NMR spectrometer
(together with some post-processing) outputs the distances between pairs of
atoms in a large molecule. The distances are not specified for all atom pairs,
and they are uncertain---given only up to an interval. This setup lends itself
naturally to EDM treatment; for example, it can be directly addressed using
MDS \cite{trosset1998}. Indeed, the crystallography community also contributed
a large number of important results on distance geometry. In a different
biochemical application, comparing distance matrices yields efficient
algorithms for comparing proteins from their 3D structure
\cite{Holm:1993dx}.

In machine learning, one can learn manifolds by finding an EDM with a low
embedding dimension that preserves the geometry of local neighborhoods.
Weinberger and Saul use it to learn image manifolds
\cite{Weinberger2004}. Other examples of using Euclidean distance geometry in
machine learning are results by Tenenbaum, De Silva and Langford
\cite{Tenenbaum2000} on image understanding and handwriting recognition, Jain
and Saul \cite{jain2004} on speech and music, and Demaine and \emph{et al.}
\cite{Demaine:2009dw} on music and musical rhythms.

With the increased interest in sensor networks, several EDM-based approaches
were proposed for sensor localization
\cite{Alfakih1999,dohetry2001,Biswas2004,Dattorro:2011wa}. 
Connections between EDMs, multilateration and semidefinite programming are
expounded in depth in \cite{So:2007cz}, especially in the context of sensor
network localization.

Position calibration in ad-hoc microphone arrays is often done with sources at
unknown locations, such as handclaps, fingersnaps or randomly placed
loudspeakers
\cite{Crocco:2012eu,Pollefeys:2008ho,Dokmanic:2014tc}. This gives us distances
(possibly up to an offset time) between the microphones and the sources and
leads to the problem of multi-dimensional unfolding
\cite{Schonemann:1970wd}.

All of the above applications work with labeled distance data. In certain TOA-
based applications one loses the labels---the correct permutation of the
distances is no longer known. This arises in reconstructing the geometry of a
room from echoes
\cite{Dokmanic:2013dz}.
Another example of unlabeled distances is in sparse phase retrieval, where the
distances between the unknown non-zero lags in a signal are revealed in its
autocorrelation function \cite{Ranieri:2013tx}. \rev{Recently, motivated by
problems in crystallography, Gujarahati and co-%
authors published an algorithm for reconstruction of Euclidean networks from
unlabeled distance data \cite{Gujarathi:2014cz}.}

\subsection{Our Mission}

We were motivated to write this tutorial after realizing that EDMs are not
common knowledge in the signal processing community, perhaps for the lack of a
compact introductory text. This is effectively illustrated by the anecdote
that, not long before writing this article, one of the authors had to add the
(rather fundamental) rank property to the Wikipedia page on EDMs.\footnote{We
are working on improving that page substantially.} In a compact tutorial we do
not attempt to be exhaustive; much more thorough literature reviews are
available in longer expos\'es on EDMs and distance geometry
\cite{Liberti:2012ut,Krislock:2012xx,Mucherino:2012hw}. Unlike these works
that take the most general approach through graph realizations, we opt to show
simple cases through examples, and to explain and spell out a set of basic
algorithms that anyone can use immediately. Two big topics that we discuss are
not commonly treated in the EDM literature: localization from unlabeled
distances, and multidimensional unfolding (applied to microphone
localization). On the other hand, we choose to not explicitly discuss the
sensor network localization (SNL) problem, as the relevant literature is
abundant.

Implementations of all the algorithms are available
online.\footnote{\url{http://lcav.epfl.ch/ivan.dokmanic}} Our hope is that
this will provide a good entry point for those wishing to learn much more, and
inspire new approaches to old problems.

\begin{table}[h!]
  \centering
  \caption{Summary of notation}
  \begin{tabular}{@{}p{0.5in}p{2.7in}@{}}
  \toprule
  {\bf Symbol} & {\bf Meaning} \\
  \midrule
  $n$ &
  Number of points (columns) in $\mX = [\vx_1, \ldots, \vx_n]$\\
  $d$ &
  Dimensionality of the Euclidean space\\
  $a_{ij}$ &
  Element of a matrix $\mA$ on the $i$th row and the $j$th column\\
  $\mD$ &
  A Euclidean distance matrix \\
  $\EDM(\mX)$ &
  Euclidean distance matrix created from columns in $\mX$ \\ 
  $\EDM(\mX,\mY)$ &
  Matrix containing the squared distances between the columns of $\mX$ and $\mY$ \\
  $\EDMgram(\mG)$ &
  Euclidean distance matrix created from the Gram matrix $\mG$ \\
  $\mJ$ &
  Geometric centering matrix \\
  $\mA_{\mW}$ &
  Restriction of $\mA$ to non-zero entries in $\mW$ \\
  $\mW$ &
  Mask matrix, with ones for observed entries \\
  $\S_+^n$ &
  Set of real symmetric positive semidefinite matrices in $\R^{n \times n}$ \\
  $\affdim(\mX)$ &
  Affine dimension of the points listed in $\mX$ \\
  $\mA \circ \mB$ &
  Hadamard (entrywise) product of $\mA$ and $\mB$ \\
  $\epsilon_{ij}$ &
  Noise corrupting the $(i,j)$ distance\\
  $\ve_i$ &
  $i$th vector of the canonical basis\\
  $\norm{ \mA }_F$ &
  Frobenius norm of $\mA$, $\big(\sum_{ij} a_{ij}^2\big)^{1/2}$\\
  \bottomrule
  \end{tabular}
  \label{tab:notation-summary}
\end{table}

\section{From Points to EDMs and Back} % (fold)
\label{sec:from_points_to_edms_and_back}
  
The principal EDM-related task is to reconstruct the original point set. This
task is an inverse problem to the simpler forward problem of finding the EDM
given the points. Thus it is desirable to have an analytic expression for the
EDM in terms of the point matrix. Beyond convenience, we can expect such an
expression to provide interesting structural insights. We will define notation
as it becomes necessary---a summary is given in Table
\ref{tab:notation-summary}.

Consider a collection of $n$ points in a $d$-dimensional Euclidean space,
ascribed to the columns of matrix $\mX \in \R^{d \times n}$, $\mX = [\vx_1, \
\vx_2, \ \cdots,\
\vx_n], \ \vx_i \in \R^d$. Then the squared distance between $
\vx_i$ and $\vx_j$ is given as
\begin{equation}
	d_{ij} = \norm{\vx_i - \vx_j}^2,
\end{equation}
where $\norm{\, \cdot \,}$ denotes the Euclidean norm. Expanding the norm
yields
\begin{equation}
	d_{ij} = (\vx_i-\vx_j)^\T(\vx_i-\vx_j) = \vx_i^\T \vx_i - 2\vx_i^\T \vx_j + \vx_j^\T \vx_j.
\end{equation}
From here, we can read out the matrix equation for $\mD = [d_{ij}]$,
\begin{equation}
	\label{eq:edm_assemble}
	\begin{aligned}
	\EDM(\mX) \bydef \vone \diag(\mX^\T \mX)^\T - 2\mX^\T \mX + \diag(\mX^\T
	\mX) \vone^\T,
	\end{aligned}
\end{equation}
where $\vone$ denotes the column vector of all ones and $\diag(\mA)$ is a
column vector of the diagonal entries of $\mA$. We see that $\EDM(\mX)$ is in
fact a function of $\mX^\T \mX$. For later reference, it is convenient to
define an operator $\EDMgram(\mG)$ similar to $\EDM(\mX)$, that operates
directly on the Gram matrix $\mG = \mX^\T \mX$,
\begin{equation}
	\label{eq:edm_gram_assemble}
	\EDMgram(\mG) \bydef \diag(\mG) \vone^\T  - 2 \mG + \vone \diag(\mG)^\T.
\end{equation}

The EDM assembly formula \eqref{eq:edm_assemble} or
\eqref{eq:edm_gram_assemble} reveals an important property: Because the rank
of $\mX$ is at most $d$ (it has $d$ rows), then the rank of $\mX^\T \mX$ is
also at most $d$. The remaining two summands in
\eqref{eq:edm_assemble} have rank one. By rank inequalities, rank of a sum of
matrices cannot exceed the sum of the ranks of the summands. With this
observation, we proved one of the most notable facts about EDMs:

\begin{thm}[Rank of EDMs]
\label{thm:edm_rank}
Rank of an EDM corresponding to points in $\R^d$ is at most $d+2$.
\end{thm}

This is a powerful theorem: it states that the rank of an EDM is independent
of the number of points that generate it. In many applications, $d$ is three
or less, while $n$ can be in the thousands. According to Theorem
\ref{thm:edm_rank}, rank of such practical matrices is at most five. The proof
of this theorem is simple, but to appreciate that the property is not obvious,
you may try to compute the rank of the matrix of non-squared distances.

\begin{figure}
\centering
\includegraphics[width=3.5in]{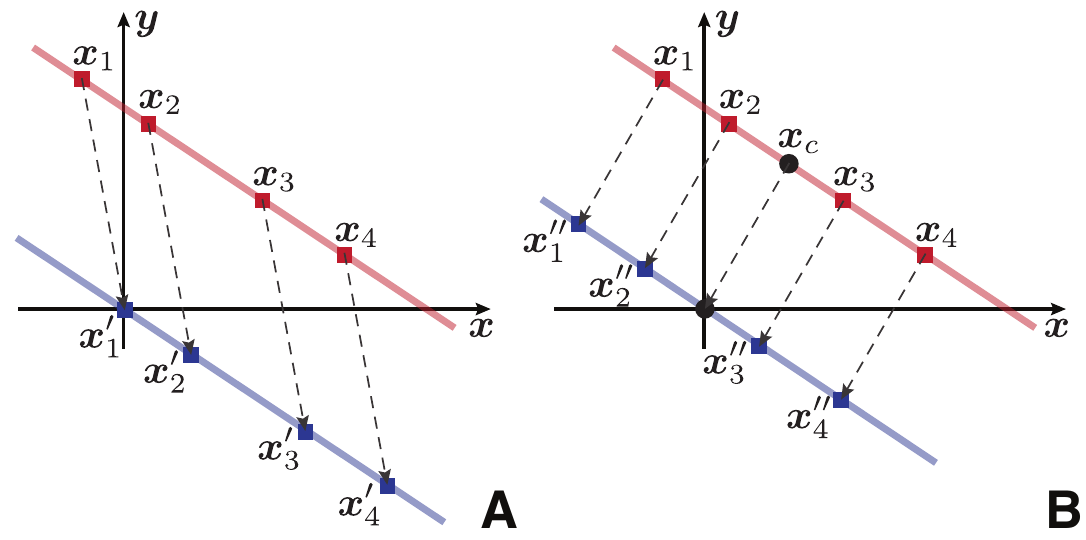}
\caption{Illustration of the relationship between an affine subspace and its
parallel linear subspace. The points $\mX = [\vx_1, \ldots, \vx_4]$ live in an
affine subspace---a line in $\R^2$ that does not contain the origin. In (A),
the vector $\vx_1$ is subtracted from all the points, and the new point list
is $\mX' = [\vzero, \vx_2 - \vx_1, \vx_3 - \vx_1, \vx_4 - \vx_1]$. While the
columns of $\mX$ span $\R^2$, the columns of $\mX'$ only span the 1D subspace
of $\R^2$---the line through the origin. In (B), we subtract a different
vector from all points: the centroid $\frac{1}{4} \mX
\vone$. The translated vectors $\mX'' = [\vx_1'', \ldots, \vx_4'']$ again span
the same 1D subspace.}
\label{fig:affine}
\end{figure}

What really matters in Theorem \ref{thm:edm_rank} is the affine dimension of
the point set---the dimension of the smallest affine subspace that contains
the points, denoted by $\affdim(\mX)$. For example, if the points lie on a
plane (but not on a line or a circle) in $\R^3$, rank of the corresponding EDM
is four, not five. This will be clear from a different perspective in the next
subsection, as any affine subspace is just a translation of a linear subspace.
An illustration for a 1D subspace of $\R^2$ is provided in Fig.
\ref{fig:affine}: Subtracting \emph{any} point in the affine subspace from all
its points translates it to the parallel linear subspace that contains the
zero vector.

\subsection{Essential Uniqueness} % (fold)
\label{sub:essential_uniqueness}

When solving an inverse problem, we need to understand what is recoverable and
what is forever lost in the forward problem. Representing sets of points by
distances usually increases the size of the representation. For most
interesting $n$ and $d$, the number of pairwise distances is larger than the
size of the coordinate description, $\binom{n}{2} > nd$, so an EDM holds
more scalars than the list of point coordinates. Nevertheless, some
information is lost in this encoding, namely the information about the
absolute position and orientation of the point set. Intuitively, it is clear
that rigid transformations (including reflections) do not change distances
between the fixed points in a point set. This intuitive fact is easily deduced
from the EDM assembly formula \eqref{eq:edm_assemble}. We have seen in
\eqref{eq:edm_assemble} and \eqref{eq:edm_gram_assemble} that $\EDM(\mX)$ is
in fact a function of the Gram matrix $\mX^\T \mX$.

\begin{figure}[t]
\centering
\includegraphics[width=3.3in]{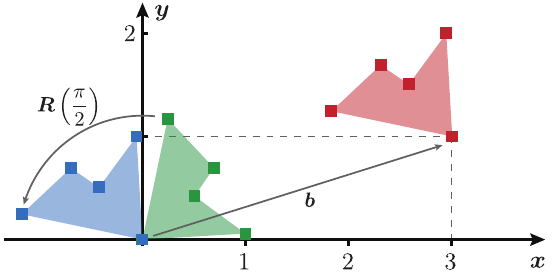}
\caption{Illustration of a rigid transformation in 2D. Here the points set is
transformed as $\mR \mX + \vb \vone^{\! \T}$. Rotation matrix \mtx,
corresponds to a counterclockwise rotation of 90$^\circ$. The translation
vector is $\vb = [3, \ 1]^\T$. The shape is drawn for visual reference.}
\label{fig:procrustes}
\end{figure}

This makes it easy to show algebraically that rotations and
reflections do not alter the distances. Any rotation/reflection can be
represented by an orthogonal matrix $\mQ \in \R^{d \times d}$ acting on the
points $\vx_i$. Thus for the rotated point set $\mX_r =
\mQ \mX$ we can write

\begin{equation}
	\mX_r^\T \mX_r = (\mQ \mX)^\T (\mQ \mX) = \mX^\T \mQ^\T \mQ \mX = \mX^\T \mX,
\end{equation}
where we invoked the orthogonality of the rotation/reflection matrix,
$\mQ^\T \mQ = \mI$.

Translation by a vector $\vb \in \R^d$ can be expressed as
\begin{equation}
	\mX_t = \mX + \vb \vone^\T.
\end{equation}
Using $\diag(\mX_t^\T \mX_t) = \diag(\mX^\T \mX) + 2 \mX^\T \vb + \norm{b}^2
\vone$, one can directly verify that this transformation leaves
\eqref{eq:edm_assemble} intact. In summary,
\begin{equation}
	\EDM(\mQ \mX) = \EDM(\mX + \vb \vone^\T) = \EDM(\mX).
\end{equation}

The consequence of this invariance is that we will never be able to
reconstruct the absolute orientation of the point set using only the
distances, and the corresponding degrees of freedom will be chosen freely.
\rev{Different reconstruction procedures will lead to different realizations of the
point set, all of them being rigid transformations of each other.} Fig.
\ref{fig:procrustes} illustrates a point set under a rigid transformation. It
is clear that the distances between the points are the same for all three
shapes.

% subsection essential_uniqueness (end)

\subsection{Reconstructing the Point Set From Distances} % (fold)
\label{sub:reconstructing_the_point_set_from_distances}

% subsection reconstructing_the_point_set_from_distances (end)
The EDM equation \eqref{eq:edm_assemble} hints at a procedure to compute the
point set starting from the distance matrix. 
%
% In the context of the discussion about uniqueness, it is clear that to do so
% we must necessarily assume a gauge.
%
Consider the following choice: let the first point $\vx_1$  be at the origin.
Then the first column of $\mD$ contains the squared norms of the point
vectors,
\begin{equation}
	d_{i1} = \norm{\vx_i - \vx_1}^2 = \norm{\vx_i - \vzero}^2 = \norm{\vx_i}^2.
\end{equation}
Consequently, we can immediately construct the term $\vone \diag(\mX^\T \mX)$
and its transpose in \eqref{eq:edm_assemble}, as the diagonal of $\mX^\T \mX$ contains exactly the norms squared $\norm{\vx_i}^2$. Concretely, 
\begin{equation}
	\vone \diag(\mX^\T \mX) = \vone \vd_1^\T,
\end{equation}
where $\vd_1 = \mD \ve_1$ is the first column of $\mD$. We thus obtain the Gram matrix
from 
\eqref{eq:edm_assemble} as
\begin{equation}
	\label{eq:classical_mds_simple}
	\mG = \mX^\T \mX = -\frac{1}{2}(\mD - \vone \vd_1^\T - \vd_1 \vone^\T).
\end{equation}
The point set can be found by an EVD, $\mG = \mU
\mLambda \mU^\T$, where $\mLambda = \diag(\lambda_1, \ldots, \lambda_n)$ with all eigenvalues
$\lambda_i$ non-negative, and $\mU$ orthonormal, as $\mG$ is a symmetric
positive semidefinite matrix. Throughout the paper we assume that the
eigenvalues are sorted in the order of decreasing magnitude, $\abs{\lambda_1}
\geq \abs{\lambda_2} \geq \cdots \geq \abs{\lambda_n}$. We can now set
$\wh{\mX} \bydef \big[ \diag \big( \sqrt{\lambda_1},\ \ldots,\
\sqrt{\lambda_d}\, \big), \ \vzero_{d \times (n-d)} \big] \mU^\T$. Note that
we could have simply taken $\mLambda^{1/2}
\mU^\T$ as the reconstructed point set, but if the Gram matrix really
describes a $d$-dimensional point set, the trailing eigenvalues will be
zeroes, so we choose to truncate the corresponding rows.

\rev{It is straightforward to verify that the reconstructed point set
$\wh{\mX}$ generates the original EDM, $\mD = \EDM(\mX)$; as we have learned,
$\wh{\mX}$ and $\mX$ are related by a rigid transformation. The described procedure
is called the \emph{classical MDS}, with a particular choice of the coordinate
system: $\vx_1$ is fixed at the origin.}

In \eqref{eq:classical_mds_simple} we subtract a structured rank-2 matrix
$(\vone \vd_1^\T + \vd_1 \vone^\T)$ from $\mD$. A more systematic approach to
the classical MDS is to use a generalization of
\eqref{eq:classical_mds_simple} by Gower
\cite{gower1982}. Any such subtraction that makes the right hand side of
\eqref{eq:classical_mds_simple} positive semidefinite (PSD), \emph{i.e.}, that
makes $\mG$ a Gram matrix, can also be modeled by multiplying $\mD$ from both
sides by a particular matrix. This is substantiated in the following result.

\begin{thm}[Gower \cite{gower1982}]
\label{thm:sdp_equivalence}
$\mD$ is an EDM if and only if
\begin{equation}
	\label{eq:sdp_equivalence}
    -\frac{1}{2}(\mI - \vone \vs^\T) \mD (\mI  - \vs \vone^\T)
\end{equation}
is PSD for any $\vs$ such that $\vs^\T \vone =
1$ and $\vs^\T \mD \neq \vzero$.
\end{thm}
In fact, if \eqref{eq:sdp_equivalence} is PSD for one such $\vs$, then it is
PSD for all of them. In particular, define the \emph{geometric centering
matrix} as
\begin{equation}
	\mJ \bydef \mI - \frac{1}{n} \vone \vone^\T.
\end{equation}
Then $-\frac{1}{2} \mJ \mD \mJ$ being positive semidefinite is equivalent to
$\mD$ being an EDM. Different choices of $\vs$ correspond to different
translations of the point set. 

The classical MDS algorithm with the geometric centering matrix is spelled out
in Algorithm \ref{alg:classical_mds}. \rev{Whereas so far we have assumed that
the distance measurements are noiseless, Algorithm \ref{alg:classical_mds} can
handle noisy distances too, as it discards all but the $d$ largest
eigenvalues.}

\begin{algorithm}[t]
\caption{Classical MDS}
\label{alg:classical_mds}
\begin{algorithmic}[1]
\Function{ClassicalMDS}{$\mD, d$}
	\State $\mJ \gets \mI - \frac{1}{n}\vone \vone^\T$ \Comment{Geometric centering matrix}
	\State $\mG \gets -\frac{1}{2} \mJ \mD \mJ$ \Comment{Compute the Gram matrix}
	% \State $\mG \gets \mathrm{SVThreshold}(\mG, d)$
	% \State $\mU, \mS, \mV \gets \mathrm{ThinSVD}(\mG)$
  \State $\mU, [\,\lambda_i\,]_{i=1}^n \gets \EVD(\mG)$
	\State \Return{$\big[ \diag \big( \sqrt{\lambda_1},\ \ldots,\ \sqrt{\lambda_d}\, \big), \ \vzero_{d \times (n-d)} \big] \mU^\T$}
\EndFunction
% \State
\end{algorithmic}
\end{algorithm}

It is straightforward to verify that \eqref{eq:classical_mds_simple}
corresponds to $\vs = \ve_1$. Think about what this means in terms of the
point set: $\mX \ve_1$ selects the first point in the list, $\vx_1$. Then
$\mX_0 = \mX (\mI - \ve_1 \vone^\T)$ translates the points so that $\vx_1$ is
translated to the origin. Multiplying the definition \eqref{eq:edm_assemble}
from the right by $(\mI - \ve_1 \vone^\T)$ and from the left by $(\mI - \vone
\ve_1^\T)$ will annihilate the two rank-1 matrices, $\diag(\mG) \vone^\T$ and
$\vone \diag(\mG)^\T$. We see that the remaining term has the form $-2
\mX_0^\T \mX_0$, and the reconstructed point set will have the first point at
the origin!

On the other hand, setting $\vs = \frac{1}{n} \vone$ places the centroid of
the point set at the origin of the coordinate system. For this reason, the
matrix $\mJ = \mI -
\frac{1}{n}\vone \vone^\T$ is called the \emph{centering matrix}. To better
understand why, consider how we normally center a set of points given in
$\mX$. 

First, we compute the centroid as the mean of all the points
\begin{equation}
	\vx_c = \frac{1}{n}\sum_{i=1}^n \vx_i = \frac{1}{n} \mX \vone.
\end{equation}
Second, we subtract this vector from all the points in the set,
\begin{equation}
	\label{eq:centering_derivation}
	\mX_c = \mX - \vx_c \vone^\T = \mX - \frac{1}{n} \mX \vone \vone^\T = \mX (\mI
	- \frac{1}{n} \vone \vone^\T).
\end{equation}
In complete analogy with the reasoning for $\vs = \ve_1$, we can see that the
reconstructed point set will be centered at the origin.

\subsection{Orthogonal Procrustes Problem}

Since the absolute position and orientation of the points are lost when going
over to distances, we need a method to align the reconstructed point set with
a set of \emph{anchors}---points whose coordinates are fixed and known.

This can be achieved in two steps, sometimes called Procrustes analysis.
Ascribe the anchors to the columns of $\mY$, and suppose that we want to align
the point set $\mX$ with the columns of $\mY$. \rev{Let $\mX_a$ denote the
submatrix (a selection of columns) of $\mX$ that should be aligned with the
anchors. We note that the number of anchors---columns in $\mX_a$---is typically
small compared with the total number of points---columns in $\mX$.}

In the first step, we remove the means $\vy_c$ and $\vx_{a,c}$ from matrices
$\mY$ and $\mX_a$, obtaining the matrices $\overline{\mY}$ and
$\overline{\mX}_a$. In the second step, termed orthogonal Procrustes analysis,
we are searching for the rotation and reflection that best maps
$\overline{\mX}_a$ onto $\overline{\mY}$,

\begin{equation}
	\label{eq:orthogonal_procrustes}
	\mR = \argmin_{\mQ:\mQ \mQ^\T = \mI} \norm{\mQ \overline{\mX}_a - \overline{\mY}}_F^2.
\end{equation}
\rev{The Frobenius norm $\norm{\,\cdot\,}_F$ is simply the $\ell^2$-norm of
the matrix entries, $\norm{\mA}_F^2 \bydef \sum
a_{ij}^{\,2} = \trace(\mA^\T \mA)$. 

The solution to
\eqref{eq:orthogonal_procrustes}---found by Sch\"onemann in his PhD thesis
\cite{Schonemann:1964tj}---is given by the singular value decomposition (SVD).
Let $\overline{\mX}_a \overline{\mY}^\T = \mU
\mSigma \mV^\T$; then we can continue computing
\eqref{eq:orthogonal_procrustes} as follows 

\begin{align}
  \label{eq:procrustes_solution}
  \mR
  &= \argmin_{\mQ:\mQ \mQ^\T = \mI}  \norm{\mQ \overline{\mX}_a}_F^2 + \norm{\overline{\mY}}_F^2 - \trace(\mY^\T \mQ \overline{\mX}_a)  \nonumber \\
  &= \argmax_{\wt{\mQ}:\wt{\mQ} \wt{\mQ}^\T = \mI}  \trace(\wt{\mQ} \, \mSigma), 
\end{align}
where $\wt{\mQ} \bydef \mV^\T \mQ \mU$, and we used the orthogonal invariance
of the Frobenius norm and the cyclic invariance of the trace. The
last trace expression in \eqref{eq:procrustes_solution} is equal to
$\sum_{i=1}^n  \sigma_i \wt{q}_{ii}$. Noting that $\wt{\mQ}$ is also an
orthogonal matrix, its diagonal entries cannot exceed 1. Therefore, the
maximum is achieved when $\wt{q}_{ii} = 1$ for all $i$, meaning that the
optimal $\wt{\mQ}$ is an identity matrix. It follows that $\mR = \mV
\mU^\T$. 

Once the optimal rigid transformation has been found, the alignment can be
applied to the entire point set as
\begin{equation}
	\mR (\mX - \vx_{a,c} \vone^\T) + \vy_c \vone^\T.
\end{equation}
}

\subsection{Counting the Degrees of Freedom}

It is interesting to count how many degrees of freedom there are in different
EDM related objects. Clearly, for $n$ points in $\R^d$ we have
\begin{equation}
	\#_{\mX} = n \times d
\end{equation}
degrees of freedom: If we describe the point set by the list of coordinates,
the size of the description matches the number of degrees of freedom. Going
from the points to the EDM (usually) increases the description size to
$\frac{1}{2}{n(n-1)}$, as the EDM lists the distances between all the pairs of
points. 
% On the other hand, we learned that we lose information about the
% absolute orientation. 
By Theorem \ref{thm:edm_rank} we know that the EDM has
rank at most $d+2$.

\rev{Let us imagine for a moment that we do not know any other EDM-specific
properties of our matrix, except that it is symmetric, positive, zero-diagonal
(or \emph{hollow}), and that it has rank $d+2$. The purpose of this exercise
is to count the degrees of freedom associated with such a matrix, and to see
if their number matches the intrinsic number of the degrees of freedom of the
point set, $\#_{\mX}$. If it did, then these properties would completely
characterize an EDM. We can already anticipate from Theorem
\ref{thm:sdp_equivalence} that we need more properties: a certain matrix related to the
EDM---as given in \eqref{eq:sdp_equivalence}---must be PSD. Still, we want to
see how many degrees of freedom we miss.

We can do the counting by looking at the EVD of a symmetric matrix, $\mD =
\mU \mLambda \mU^\T$. The diagonal matrix $\mLambda$ is specified by $d+2$ degrees
of freedom, because $\mD$ has rank $d+2$. The first eigenvector of length $n$
takes up $n-1$ degrees of freedom due to the normalization; the second one
takes up $n-2$, as it is in addition orthogonal to the first one; for the last
eigenvector, number $(d+2)$, we need $n - (d+2)$ degrees of freedom. We do not
need to count the other eigenvectors, because they correspond to zero
eigenvalues. The total number is then

\begin{align*}
  \#_\text{DOF} &= \underbrace{(d+2)}_{\text{Eigenvalues}} + \underbrace{(n-1) + \cdots + [n-(d+2)]}_{\text{Eigenvectors}} \ - \underbrace{\phantom{(}n\phantom{)}}_{\mathclap{\text{Hollowness}}} \nonumber \\
	 &= n \times (d+1) - \frac{(d+1) \times (d+2)}{2} \nonumber \\
\end{align*}
}

For large $n$ and fixed $d$, it follows that
\begin{equation}
	\label{eq:dof_ratio}
	\frac{\#_\text{DOF}}{\#_{\mX}} \sim \frac{d+1}{d}.
\end{equation}
Therefore, even though the rank property is useful and we will show efficient
algorithms that exploit it, it is still not a \emph{tight} property (symmetry
and hollowness included). For $d=3$, the ratio
\eqref{eq:dof_ratio} is $\frac{4}{3}$, so loosely speaking, the rank property
has 30\% determining scalars too many, which we need to set consistently.
Put differently, we need 30\% more data in order to exploit the rank property
than we need to exploit the full EDM structure. Again loosely phrased, we can
assert that for the same amount of data, the algorithms perform at least
$\approx$30\% worse if we only exploit the rank property, without
\emph{EDMness}.

The one-third gap accounts for various geometrical constraints that must be
satisfied. The redundancy in the EDM representation is what makes denoising
and completion algorithms possible, and thinking in terms of degrees of freedom
gives us a fundamental understanding of what is achievable. Interestingly, the
above discussion suggests that for large $n$ and large $d = o(n)$, little is
lost by only considering rank.

\rev{Finally, in the above discussion, for the sake of simplicity we ignored the
degrees of freedom related to absolute orientation. These degrees of freedom,
not present in the EDM, do not affect the large-$n$ behavior.}

\subsection{Summary}

Let us summarize what we have achieved in this section:
\begin{itemize}
	\item We explained how to algebraically construct an EDM given the list of
	point coordinates,
	\item We discussed the essential uniqueness of the point set; information
	about the absolute orientation of the points is irretrievably lost when
	transitioning from points to an EDM,
  \item We explained classical MDS---a simple eigenvalue-decomposition-based algorithm
  (Algorithm \ref{alg:classical_mds}) for reconstructing the original points%
  ---along with discussing parameter choices that lead to different centroids in
  reconstruction,
	\item Degrees-of-freedom provide insight into scaling behavior. We
	showed that the rank property is pretty good, but there is more to it than
	just rank.
\end{itemize}

% section from_points_to_edms_and_back (end)

\section{EDMs as a Practical Tool} % (fold)
\label{sec:edms_as_a_practical_tool}

We rarely have a perfect EDM. Not only are the entries of the measured matrix
plagued by errors, but often we can measure just a subset. There are various
sources of error in distance measurements: we already know that in NMR
spectroscopy, instead of exact distances we get intervals. Measuring distance
using received powers or TOAs is subject to noise, sampling errors and model
mismatch.

Missing entries arise because of the limited radio range, or because of the
nature of the spectrometer. Sometimes the nodes in the problem at hand are
asymmetric by definition; in microphone calibration we have two types:
microphones and calibration sources. This results in a particular block
structure of the missing entries (we will come back to this later, but you can
fast-forward to Fig.~\ref{fig:mdu} for an illustration).

It is convenient to have a single statement for both EDM approximation and EDM
completion, as the algorithms described in this section handle them at once.

\begin{problem}
\label{prob:approximation_completion}
Let $\mD = \EDM(\mX)$. We are given a noisy observation of the distances
between $p \leq \binom{n}{2}$ pairs of points from $\mX$. That is, we have a
noisy measurement of $2p$ entries in $\mD$,
\begin{equation}
	\wt{d}_{ij} = d_{ij} + \epsilon_{ij},
\end{equation}
for $(i, j) \in E$, where $E$ is some index set, and $\epsilon_{ij}$ absorbs
all errors. The goal is to reconstruct the point set $\wh{\mX}$ in the given
embedding dimension, so that the entries of $\EDM(\wh{\mX})$ are close in some
metric to the observed entries $\wt{d}_{ij}$.
\end{problem}

To concisely write down completion problems, we define the mask matrix $\mW$
as follows,
\begin{equation}
	w_{ij} \bydef
	\begin{cases}
	   1, \ (i, j) \in E \\
	   0, \ \text{otherwise}.
	\end{cases}
\end{equation}
This matrix then selects elements of an EDM through a Hadamard (entrywise)
product. For example, to compute the norm of the difference between the
observed entries in $\mA$ and $\mB$, we write $\norm{\mW \circ (\mA - \mB)}$.
Furthermore, we define the indexing $\mA_{\mW}$ to mean the restriction of
$\mA$ to those entries where $\mW$ is non-zero. The meaning of $\mB_{\mW}
\gets \mA_{\mW}$ is that we assign the observed part of $\mA$ to the observed
part of $\mB$.

\subsection{Exploiting the Rank Property} % (fold)
\label{sub:exploiting_the_rank_property}

Perhaps the most notable fact about EDMs is the rank property established in
Theorem \ref{thm:edm_rank}: The rank of an EDM for points living in $\R^d$ is
at most $d+2$. This leads to conceptually simple algorithms for EDM completion
and denoising. Interestingly, these algorithms exploit only the rank of the
EDM. There is no explicit Euclidean geometry involved, at least not before
reconstructing the point set.

We have two pieces of information: a subset of potentially noisy distances,
and the desired embedding dimension of the point configuration. The latter
implies the rank property of the EDM that we aim to exploit. We may try to
alternate between enforcing these two properties, and hope that the algorithm
produces a sequence of matrices that converges to an EDM. If it does, we have
a solution. Alternatively, it may happen that we converge to a matrix with the
correct rank that is not an EDM, or that the algorithm never converges. The
pseudocode is listed in Algorithm \ref{alg:rank_complete_edm}.

\begin{algorithm}[t]
\caption{Alternating Rank-Based EDM Completion}
\label{alg:rank_complete_edm}
\begin{algorithmic}[1]
\Function{RankCompleteEDM}{$\mW, \wt{\mD}, d$}

	\State $\mD_{\mW} \gets \wt{\mD}_{\mW}$ \Comment{Initialize observed entries}
	\State $\mD_{\vone\vone^\T - \mW} \gets \mu$ \Comment{Initialize unobserved entries}
	\Repeat
			\State $\mD \gets \mathrm{EVThreshold}(\mD,\, d+2)$
			\State $\mD_{\mW} \gets \wt{\mD}_{\mW}$ \Comment{Enforce known entries}
			\State $\mD_{\mI} \gets \vzero$ \Comment{Set the diagonal to zero}
			\State $\mD \gets (\mD)_+$ \Comment{Zero the negative entries}
	\Until{Convergence or MaxIter}
	\State \Return{\mD}
\EndFunction
\break \raggedright
\Function{EVThreshold}{$\mD, r$}
  \State $\mU, [\lambda_i]_{i=1}^n \gets \EVD(\mD)$
  \State $\mSigma \gets \diag \big(\lambda_1, \ldots, \lambda_r, \underbrace{0, \ldots, 0}_{n-r~\text{times}}\big)$
  \vspace{-3mm}
  \State $\mD \gets \mU \mSigma \mU^\T$
  \State \Return{$\mD$}
\EndFunction
\end{algorithmic}
\end{algorithm}

A different, more powerful approach is to leverage algorithms for low rank
matrix completion developed by the compressed sensing community. For example,
OptSpace \cite{Keshavan:2010bt} is an algorithm for recovering a low-rank
matrix from noisy, incomplete data. \rev{Let us take a look at how OptSpace works.
Denote by $\mM \in \R^{m \times n}$ the rank-$r$ matrix that we seek to recover,
by $\mZ \in \R^{m \times n}$ the measurement noise, and by $\mW \in \R^{m
\times n}$ the mask corresponding to the measured entries; for simplicity we
choose $m \leq n$. The measured noisy and incomplete matrix is then given as
\begin{equation}
	\wt{\mM} = \mW \circ (\mM + \mZ).
\end{equation}
Effectively, this sets the missing (non-observed) entries of the matrix to
zero. OptSpace aims to minimize the following cost function,
\begin{equation}
  \label{eq:optspace_cost_function}
  F(\mA, \mS, \mB) \bydef  \frac{1}{2} \norm{\mW
  \circ (\wt{\mM} - \mA \mS \mB^\T)}_F^2,
\end{equation}
where $\mS \in \R^{r \times r}$, $\mA \in \R^{m \times r}$, and $\mB \in \R^{n
\times r}$ such that $\mA^\T\!\mA = \mB^\T\!\mB = \mI$. Note that $\mS$ need
not be diagonal.

The cost function \eqref{eq:optspace_cost_function} is not convex, and
minimizing it is \emph{a priori} difficult \cite{Keshavan:2012tb} due to many
local minima. Nevertheless, Keshavan, Montanari and Oh \cite{Keshavan:2010bt}
show that using the gradient descent method to solve
\eqref{eq:optspace_cost_function} yields the global optimum with high
probability, provided that the descent is correctly initialized.

Let $\wt{\mM} = \sum_{i=1}^m \sigma_i \va_i \vb_i^\T$ be the SVD of
$\wt{\mM}$. Then we define the scaled rank-$r$ projection of $\wt{\mM}$ as
$\wt{\mM}_r \bydef \alpha^{-1} \sum_{i=1}^r \sigma_i
\va_i \vb_i^\T$. The fraction of observed entries is denoted by $\alpha$, so
that the scaling factor compensates the smaller \emph{average} magnitude of
the entries in $\wt{\mM}$ in comparison with $\mM$. The SVD of $\wt{\mM}_r$ is
then used to initialize the gradient descent, as detailed in Algorithm
\ref{alg:optspace}.

Two additional remarks are due in the description of OptSpace. First, it can
be shown that the performance is improved by zeroing the \emph{over-%
represented} rows and columns. A row (\emph{resp}. column) is over-represented
if it contains more than twice the average number of observed entries per row
(\emph{resp.} column). These heavy rows and columns bias the corresponding
singular vectors and singular values, so (perhaps surprisingly) it is better
to throw them away. We call this step ``Trim'' in Algorithm
\ref{alg:optspace}.

Second, the minimization of \eqref{eq:optspace_cost_function} does not have to
be performed for all variables at once. In \cite{Keshavan:2010bt}, the authors
first solve the easier, convex minimization for $\mS$, and then with the
optimizer $\mS$ fixed, they find the matrices $\mA$ and $\mB$ using the
gradient descent. These steps correspond to lines 6 and 7 of Algorithm
\ref{alg:optspace}.} For an application of OptSpace in calibration of
ultrasound measurement rigs, see the ``\textbf{Calibration in Ultrasound
Tomography}'' box.

\begin{algorithm}[t]
\caption{{\sc OptSpace} \cite{Keshavan:2010bt}} 
\label{alg:optspace} 
\begin{algorithmic}[1]
\Function{OptSpace}{$\wt{\mM}, r$}
	\State $\wt{\mM} \gets \mathrm{Trim}(\wt{\mM})$
	% \State $\mM_2 \gets \mathrm{SVThreshold}(\mM_1, r)$
	\State $\wt{\mA}, \wt{\mSigma}, \wt{\mB} \gets \SVD(\alpha^{-1} \wt{\mM})$
  % \State $\mS_0 \gets \diag(\sigma_1,\ \ldots,\ \sigma_r)$
  \State $\mA_0 \gets$ First $r$ columns of $\wt{\mA}$
  \State $\mB_0 \gets$ First $r$ columns of $\wt{\mB}$
  \State $\mS_0 \gets \displaystyle \argmin_{\mS \in \R^{r \times r}} F(\mA_0, \mS, \mB_0)$ \Comment{Eq. \eqref{eq:optspace_cost_function}}
	\State $\mA, \mB \gets \displaystyle \argmin_{\mathclap{\mA^\T\!\mA = \mB^\T\!\mB = \mI}} F(\mA, \mS_0, \mB)$ \Comment{See the note below}
  \vspace{1mm}
	\State \Return{$\mA \mS_0 \mB^\T$}
\EndFunction
\end{algorithmic}
$\rhd$ Line 7: gradient descent starting at $\mA_0, \mB_0$
\end{algorithm}

% subsection exploiting_the_rank_property (end)

\begin{figure*}[t]
\fontfamily{lmss}\selectfont
\begin{spmagbox}
\textbf{Calibration in Ultrasound Tomography}
\\
\label{sub:calibration}
\fontsize{9pt}{10pt}\selectfont

\begin{multicols}{2}

The rank property of EDMs, introduced in Theorem \ref{thm:edm_rank} can be
leveraged in calibration of ultrasound tomography devices. An example device
for diagnosing breast cancer is a circular ring with thousands of ultrasound
transducers, placed around the breast \cite{dur07}. The setup is shown in Fig.
\ref{fig:ultrasound}\textsl{A}. 

Due to manufacturing errors, the sensors are not located on a perfect circle.
This uncertainty in the positions of the sensors negatively affects the
algorithms for imaging the breast. Fortunately, we can use the measured
distances between the sensors to calibrate their relative positions. We can
estimate the distances by measuring the times-of-flight (TOFs) between pairs
of transducers in a homogeneous environment, \emph{e.g.} in water.

We cannot estimate the distances between all pairs of sensors because the
sensors have limited beam widths (it is hard to manufacture omni-directional
ultrasonic sensors). Therefore, the distances between neighboring sensors are
unknown, contrary to typical SNL scenarios where only the distances between
nearby nodes can be measured. Moreover, the distances are noisy and some of
them are unreliably estimated. This yields a noisy and incomplete EDM, whose
structure is illustrated in Figure
\ref{fig:ultrasound}\textsl{B}. 

Assuming that the sensors lie in the same plane, the original EDM produced by
them would have a rank less than five. We can use the rank property and a
low-rank matrix completion method, such as OptSpace (Algorithm
\ref{alg:optspace}), to complete and denoise the measured matrix
\cite{parhizkar:2013a}. Then we can use the classical MDS in Algorithm
\ref{alg:classical_mds} to estimate the relative locations of the ultrasound
sensors. 

For reasons mentioned above, SNL-specific algorithms are suboptimal when
applied to ultrasound calibration. An algorithm based on the rank property
effectively solves the problem, and enables one to derive upper bounds on the
performance error calibration mechanism, with respect to the number of sensors
and the measurement noise. The authors in \cite{parhizkar:2013a} show that the
error vanishes as the number of sensors increases.

\includegraphics[width=3.1in]{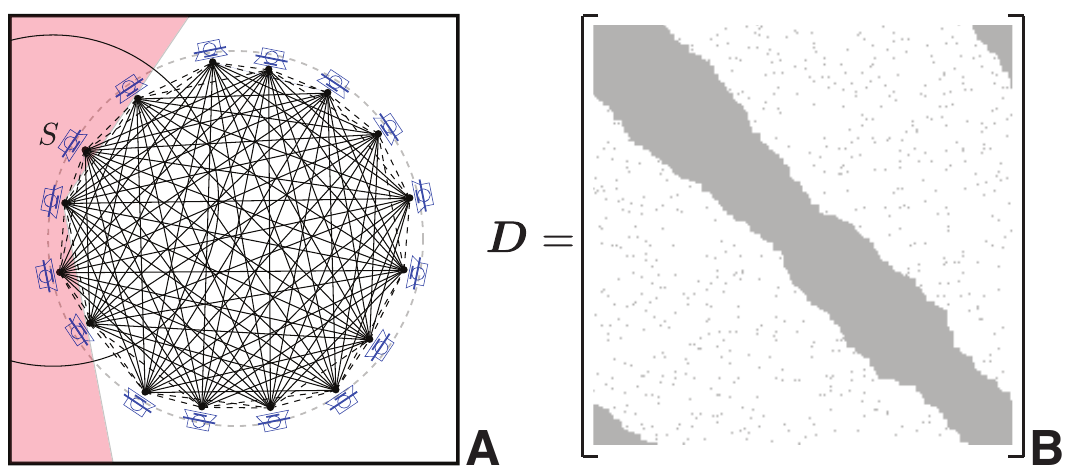}
\caption{\fontfamily{lmss}\selectfont(A) Ultrasound transducers lie on an approximately circular ring. The ring surrounds the breast and after each transducer fires an ultrasonic signal, the sound speed distribution of the breast is estimated. A precise knowledge of the sensor locations is needed to have an accurate reconstruction of the enclosed medium. (B) Because of the limited beam width of the transducers, noise and imperfect TOF estimation methods, the measured EDM is incomplete and noisy. Gray areas show missing entries of the matrix.}
\label{fig:ultrasound}

\end{multicols}
\end{spmagbox}
\end{figure*}

\subsection{Multidimensional Scaling} % (fold)
\label{sub:multidimensional_scaling}

Multidimensional scaling refers to a group of techniques that, given a set of
noisy distances, find the best fitting point conformation. It was originally
proposed in psychometrics \cite{kruskal1964,torgerson1952} to visualize the
(dis-)similarities between objects. Initially, MDS was defined as the \emph{problem}
of representing distance data, but now the term is commonly used to refer to
\emph{methods} for solving the problem \cite{Borg2005}.

Various cost functions were proposed for solving MDS. In Section
\ref{sub:reconstructing_the_point_set_from_distances}, we already encountered
one method: the classical MDS. This method minimizes the Frobenius norm of the
difference between the input matrix and the Gram matrix of the points in the
target embedding dimension. 

The Gram matrix contains inner products; rather than with inner products, it
is better to directly work with the distances. A typical cost function
represents the dissimilarity of the observed distances and the distances
between the estimated point locations. An essential observation is that the feasible
set for these optimizations is not convex (EDMs with embedding dimensions
smaller than $n-1$ lie on the boundary of a cone \cite{Dattorro:2011wa}, which
is a non-convex set).

A popular dissimilarity measure is \emph{raw stress}
\cite{kruskal1964b}, defined as the value of 
\begin{equation}
\label{eq:stress}
\begin{aligned}
& \underset{\mX\in\R^{d\times n}}{\text{minimize}} 
& & \sum_{(i, j) \in E}  \left(\sqrt{\EDM(\mX)_{ij}} - \sqrt{\wt{d}_{ij}}\right)^2\,,
\end{aligned}
\end{equation}
where $E$ defines the set of revealed elements of the distance matrix $\mD$.
The objective function can be concisely written as $\big\|\mW \circ
\big(\sqrt{\EDM(\mX)} -
\sqrt{\wt{\mD}}\big)\big\|_F^2$; a drawback of this cost function is that it
is not globally differentiable. Approaches described in the
literature comprise iterative majorization \cite{DeLeeuw1977}, various methods
using convex analysis \cite{mathar1991} and steepest descent methods
\cite{guttman1968}.

Another well-known cost function is \emph{s-stress},
\begin{equation}
\label{eq:s-stress}
\begin{aligned}
& \underset{\mX\in\R^{d\times n}}{\text{minimize}} 
& & \sum_{(i, j) \in E}  \left(\EDM(\mX)_{ij} - \wt{d}_{ij}\right)^2 \, .
\end{aligned}
\end{equation}
Again, we write the objective concisely as $\big\|\mW \circ \big(\EDM(\mX)-
\wt{\mD} \big)\big\|_F^2$. It was first studied by Takane, Young and De Leeuw
\cite{takane1977}. Conveniently, the s-stress objective is everywhere
differentiable, but at a disadvantage that it favors large over small
distances. Gaffke and Mathar \cite{gaffke1989} propose an algorithm to find
the global minimum of the s-stress function for embedding dimension $d=n-1$.
EDMs with this embedding dimension exceptionally constitute a convex set
\cite{Dattorro:2011wa}, but we are typically interested in embedding
dimensions much smaller than $n$. The s-stress minimization in
\eqref{eq:s-stress} is not convex for $d < n-1$. It was analytically shown to
have saddle points \cite{rezathesis13}, but interestingly, no analytical
non-global minimizer has been found \cite{rezathesis13}.

Browne proposed a method for computing s-stress based on Newton-Raphson root
finding \cite{Browne1987}. Glunt reports that the method by Browne  converges
to the global minimum of \eqref{eq:s-stress} in 90\% of the test cases in his
dataset\footnote{While the experimental setup of Glunt
\cite{gluntEmbedding1991} is not detailed, it was mentioned that the EDMs were
produced randomly.} \cite{gluntEmbedding1991}.

The cost function in \eqref{eq:s-stress} is separable across points $i$ and
across coordinates $k$, which is convenient for distributed implementations.
Parhizkar \cite{rezathesis13} proposed an alternating coordinate descent
method that leverages this separability, by updating a single coordinate of a
particular point at a time. The s-stress function restricted to the $k$th
coordinate of the $i$th point is a fourth-order polynomial,
\begin{equation}
\begin{aligned}
\label{eq:4th_order}
	f(x ; \bm{\alpha}^{(i,k)}) = \sum_{\ell=0}^4 \alpha_\ell^{(i,k)} x^\ell,
\end{aligned}
\end{equation}
where $\bm{\alpha}^{(i,k)}$ lists the polynomial coefficients for $i$th point
and $k$th coordinate. For example, $\alpha^{(i,k)}_0 = 4 \sum_j w_{ij}$, that
is, four times the number of points connected to point $i$. Expressions for
the remaining coefficients are given in \cite{rezathesis13}; \rev{in the
pseudocode (Algorithm
\ref{alg:alt_coor_desc}), we assume that these coefficients are returned by the
function ``GetQuadricCoeffs'', given the noisy incomplete matrix $\wt{\mD}$, the
observation mask $\mW$ and the dimensionality $d$}. The global minimizer of
\eqref{eq:4th_order} can be found analytically by calculating the roots of its
derivative (a cubic). The process is then repeated over all coordinates $k$,
and points $i$, until convergence. The resulting algorithm is remarkably
simple, yet empirically converges fast. It naturally lends itself to a
distributed implementation. We spell it out in Algorithm
\ref{alg:alt_coor_desc}.

When applied to a large dataset of random, noiseless and complete distance
matrices, Algorithm \ref{alg:alt_coor_desc} converges to the global minimum of
\eqref{eq:s-stress} in more than 99\% of the cases \cite{rezathesis13}.

\begin{algorithm}[t]
\caption{Alternating Descent \cite{rezathesis13}}
\label{alg:alt_coor_desc} 
\begin{algorithmic}[1]
\Function{AlternatingDescent}{$\wt{\mD}, \mW, d$}

	\State $\mX \in \R^{d \times n} \gets \mX_0 = \mzero$ \Comment{Initialize the point set}
	\Repeat
		\For{$i \in \{1, \cdots, n\}$} \Comment{Points}
			\For{$k \in \{1, \cdots, d\}$} \Comment{Coordinates}
				\State $\bm{\alpha}^{(i,k)} \gets \text{GetQuadricCoeffs}(\mW, \wt{\mD}, d)$
				\State $x_{i,k} \gets \argmin_{x} f(x; \bm{\alpha}^{(i, k)})$  \Comment{Eq. \eqref{eq:4th_order}}
			\EndFor
		\EndFor
	\Until{Convergence or MaxIter}
	\State \Return{\mX}
\EndFunction
\end{algorithmic}
\end{algorithm}
% subsection multidimensional_scaling (end)

\subsection{Semidefinite Programming} % (fold)
\label{sub:semidefinite_programming}

Recall the characterization of EDMs \eqref{eq:sdp_equivalence} in Theorem
\ref{thm:sdp_equivalence}. It states that $\mD$ is an EDM if and only if the
corresponding geometrically centered Gram matrix $-\frac{1}{2} \mJ \mD \mJ$ is
positive-semidefinite. Thus, it establishes a one-to-one correspondence
between the cone of EDMs, denoted by $\EDMset^n$, and the intersection of the
symmetric positive-semidefinite cone $\S_+^n$ with the geometrically centered
cone $\S_c^n$. The latter is defined as the set of all symmetric matrices
whose column sum vanishes,
\begin{equation}
	\S_c^n = \set{\mG \in \R^{n \times n} \ | \ \mG = \mG^\T,\ \mG \vone = \vzero}.
\end{equation}

We can use this correspondence to
cast EDM completion and approximation as semidefinite programs. While
\eqref{eq:sdp_equivalence} describes an EDM of an $n$-point configuration in
any dimension, we are often interested in situations where $d \ll n$. It is
easy to adjust for this case by requiring that the rank of the centered Gram
matrix be bounded. One can verify that
\begin{equation}
\label{eq:lower_dim_characterization}
\begin{rcases*}
	\mD = \EDM(\mX) \\
	\affdim(\mX)  \leq d 
\end{rcases*} \quad
\Longleftrightarrow \quad
\begin{cases}
	-\frac{1}{2} \mJ \mD \mJ \succeq 0 \\
	\rank(\mJ \mD \mJ) \leq d,
\end{cases}
\end{equation}
when $n \geq d$. That is, EDMs with a particular embedding dimension $d$ are
completely characterized by the rank and definiteness of $\mJ \mD \mJ$.

Now we can write the following rank-constrained semidefinite program for
solving Problem \ref{prob:approximation_completion},
\begin{equation}
\label{eq:rank_constrained_sdp}
\begin{aligned}
& \underset{\mG}{\text{minimize}}
& & \norm{\mW \circ \left(\wt{\mD} - \EDMgram(\mG)\right)}_F^2 \\
& \text{subject to}
& &   \rank(\mG) \leq d \\
& & & \mG \in \S^{n}_+ \cap \S^n_c.
% & & & \mG \vone = \vzero.
\end{aligned}
\end{equation}
The second constraint is just a shorthand for writing $\mG \succeq 0,\ 
\mG\vone = \vzero$. We note that this is equivalent to MDS with the s-stress
cost function, thanks to the rank characterization
\eqref{eq:lower_dim_characterization}.

Unfortunately, the rank property makes the feasible set in
\eqref{eq:rank_constrained_sdp} non-convex, and solving it exactly becomes
difficult. This makes sense, as we know that s-stress is not convex.
Nevertheless, we may \emph{relax} the hard problem, by simply omitting the
rank constraint, and hope to obtain a solution with the correct dimensionality,
\begin{equation}
\label{eq:relaxed_sdp}
\begin{aligned}
& \underset{\mG}{\text{minimize}}
& & \norm{\mW \circ \left(\wt{\mD} - \EDMgram(\mG)\right)}_F^2 \\
& \text{subject to}
& & \mG \in \S^{n}_+ \cap \S^n_c.
% & & & \mG \vone = \vzero.
\end{aligned}
\end{equation}
We call \eqref{eq:relaxed_sdp} a semidefinite relaxation (SDR) of the
rank-constrained program
\eqref{eq:rank_constrained_sdp}. 

The constraint $\mG \in \S_c^n$, or equivalently, $\mG \vone = \vzero$, means
that there are no strictly positive definite solutions ($\mG$ has a nullspace,
so at least one eigenvalue must be zero). In other words, there exist no
strictly feasible points
\cite{Krislock:2012xx}. This may pose a numerical problem, especially for
various interior point methods. The idea is then to reduce the size of the
Gram matrix through an invertible transformation, somehow removing the part of
it responsible for the nullspace. \rev{In what follows, we describe how to
construct this smaller Gram matrix.} 

A different, equivalent way to phrase the multiplicative characterization
\eqref{eq:sdp_equivalence} is the following statement: a symmetric hollow
matrix $\mD$ is an EDM if and only if it is negative semidefinite on
$\set{\vone}^\perp$ (on all vectors $\vt$ such that $\vt^\T \vone = 0$). Let
us construct an orthonormal basis for this orthogonal complement---a subspace
of dimension $(n-1)$---and arrange it in the columns of matrix $\mV \in \R^{n
\times (n-1)}$. We demand

\begin{equation}
\label{eq:V_demands}
\begin{aligned}
	&\mV^\T \vone = \vzero \\
	&\mV^\T \mV = \mI. 
\end{aligned}
\end{equation}
There are many possible choices for $\mV$, but all of them obey that
$\mV \mV^\T = \mI - \frac{1}{n}\vone
\vone^\T =  \mJ$. The following choice is given in \cite{Alfakih1999},
\begin{equation}
	\mV = 
	\begin{bmatrix}
		p      &  p     & \cdots  &  p     \\
		1+q    &  q     & \cdots  &  q     \\
		q      & 1+q    & \cdots  &  q     \\
		\vdots & \cdots & \ddots  & \vdots \\
		q      & q      & \cdots  & 1+q
	\end{bmatrix},
\end{equation}
where $p = -1/(n + \sqrt{n})$ and $q = -1 / \sqrt{n}$.

\rev{With the help of the matrix $\mV$, we can now construct the sought Gramian
with reduced dimensions.} For an EDM $\mD \in \R^{n \times n}$,
\begin{equation}
	\label{eq:gram_small}
	\GramSmall(\mD) \bydef -\frac{1}{2}\mV^\T \mD \mV
\end{equation} is an $(n-1) \times (n-1)$ PSD matrix. This
can be verified by substituting \eqref{eq:gram_small} in
\eqref{eq:edm_gram_assemble}. Additionally, we have that
\begin{equation}
\EDMgram(\mV \GramSmall(\mD) \mV^\T) = \mD.
\end{equation}
Indeed,  $\mH \mapsto \EDMgram(\mV \mH \mV^\T)$ is an invertible mapping from
$\mS_{+}^{n-1}$ to $\EDMset^n$ whose inverse is exactly $\mathcal{\mG}$. Using
these notations we can write down an equivalent optimization program that is
numerically more stable than \eqref{eq:relaxed_sdp} \cite{Alfakih1999}:

\begin{equation}
\label{eq:relaxed_sdp_schoenberg}
\begin{aligned}
& \underset{\mH}{\text{minimize}}
& & \norm{\mW \circ \left(\wt{\mD} - \EDMgram(\mV \mH \mV^\T)\right)}_F^2 \\
& \text{subject to}
& & \mH \in \S^{n-1}_+.
\end{aligned}
\end{equation}
On the one hand, with the above transformation the constraint $\mG
\vone = \vzero$ became implicit in the objective, as $\mV \mH \mV^\T \vone
\equiv \vzero$ by \eqref{eq:V_demands}; on the other hand, the feasible set is
now the full semidefinite cone $\S_+^{n-1}$.

Still, as Krislock \& Wolkowicz mention \cite{Krislock:2012xx}, by omitting
the rank constraint we allow the points to move about in a larger space, so we
may end up with a higher-dimensional solution even if there is a completion in
dimension $d$.

There exist various heuristics for promoting lower rank. One such heuristic
involves the trace norm---the convex envelope of rank. The trace or nuclear
norm is studied extensively by the compressed sensing community. In contrast
to the common wisdom in compressed sensing, the trick here is to maximize the
trace norm, not to minimize it. The mechanics are as follows: maximizing the
sum of squared distances between the points will stretch the configuration as
much as possible, subject to available constraints. But stretching favors
smaller affine dimensions (imagine pulling out a roll of paper, or stretching
a bent string). Maximizing the sum of squared distances can be rewritten as
maximizing the sum of norms in a centered point configuration---but that is
exactly the trace of the Gram matrix $\mG = -\frac{1}{2}
\mJ \mD \mJ$ \cite{Weinberger2004}. This idea has been successfully put to
work by Weinberger and Saul \cite{Weinberger2004} in manifold learning, and by
Biswas \emph{et al.} in SNL \cite{Biswas:2006cm}.

\begin{algorithm}[t]
\caption{{Semidefinite Relaxation (Matlab/CVX)}}
\vspace{-5mm}
\begin{lstlisting}
function [EDM, X] = sdr_complete_edm(D, W, lambda)

n = size(D, 1);
x = -1/(n + sqrt(n));
y = -1/sqrt(n);
V = [y*ones(1, n-1); x*ones(n-1) + eye(n-1)];
e = ones(n, 1);

cvx_begin sdp
    variable G(n-1, n-1) symmetric;
    B = V*G*V';
    E = diag(B)*e' + e*diag(B)' - 2*B;
    maximize trace(G) ...
           - lambda * norm(W .* (E - D), 'fro');
    subject to
        G >= 0;
cvx_end

[U, S, V] = svd(B);
EDM = diag(B)*e' + e*diag(B)' - 2*B;
X = sqrt(S)*V';
\end{lstlisting}
\label{alg:SDR_matlab}
\end{algorithm}

Noting that $\trace(\mH) = \trace(\mG)$ because $\trace(\mJ \mD \mJ) =
\trace(\mV^\T \mD \mV)$, we write the following SDR,
\begin{align}
\label{eq:sdp_trace_maximization}
& \underset{\mH}{\text{maximize}}
& & \trace(\mH) - \lambda \norm{\mW \circ \big(\wt{\mD} - \EDMgram (\mV \mH \mV^\T)\big)}_F \nonumber \\
& \text{subject to}
  & & \mH \in \S^{n-1}_+ 
\end{align}
Here we opted to include the data fidelity term in the Lagrangian form, as
proposed by Biswas
\cite{Biswas:2006cm}, but it could also be moved to constraints. Finally, in all of
the above relaxations, it is straightforward to include upper and lower bounds
on the distances. Because the bounds are linear constraints, the resulting
programs remain convex; this is particularly useful in the molecular
conformation problem. A Matlab/CVX \cite{cvx,gb08} implementation of the SDR
\eqref{eq:sdp_trace_maximization} is given in Algorithm \ref{alg:SDR_matlab}.

% subsection semidefinite_programming (end)

\subsection{Multidimensional Unfolding: A Special Case of Completion} % (fold)
\label{sub:box_unfolding}

\begin{figure}
\centering
\includegraphics[width=3.5in]{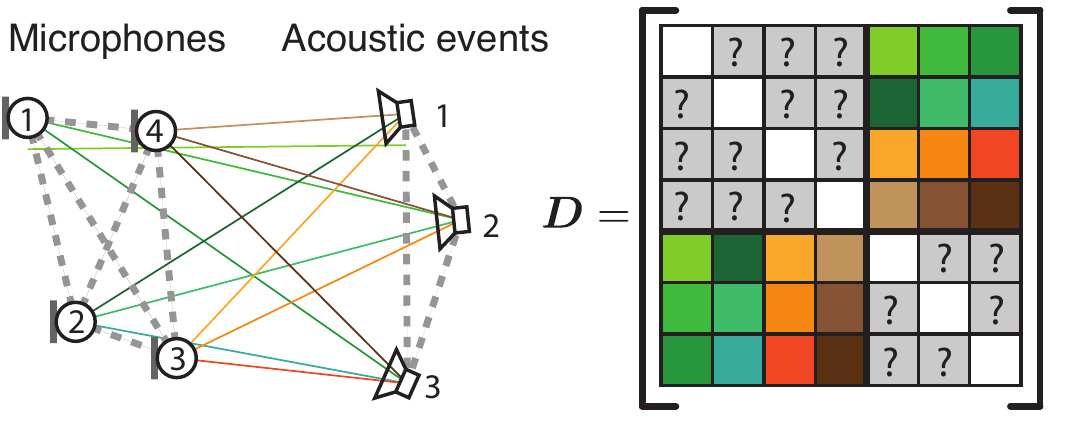}
\caption{Microphone calibration as an example of MDU. We can measure only the
propagation times from acoustic sources at unknown locations, to microphones
at unknown locations. The corresponding revealed part of the EDM has a
particular off-diagonal structure, leading to a special case of EDM
completion.}
\label{fig:mdu}
\end{figure}

Imagine that we partition the point set into two subsets, and that we can
measure the distances between the points belonging to different subsets, but
not between the points in the same subset. Metric multidimensional unfolding (MDU)
\cite{Schonemann:1970wd} refers to this special case of EDM completion. 

MDU is relevant for position calibration of ad-hoc sensor networks, in
particular of microphones. Consider an ad-hoc array of $m$ microphones at
unknown locations. We can measure the distances to $k$ point sources, also at
unknown locations, for example by emitting a pulse (we assume that the sources
and the microphones are synchronized). We can always permute the points so
that the matrix assumes the structure shown in Fig. \ref{fig:mdu}, with the
unknown entries in two diagonal blocks. This is a standard scenario described
for example in \cite{Crocco:2012eu}.

One of the early approaches to metric MDU is that of Sch\"onemann
\cite{Schonemann:1970wd}. We go through the steps of the algorithm, and then
explain how to solve the problem using the EDM toolbox. The goal is to make a
comparison, and emphasize the universality and simplicity of the introduced
tools.

\begin{figure}[t!]
  \centering
  \includegraphics[width=3.5in]{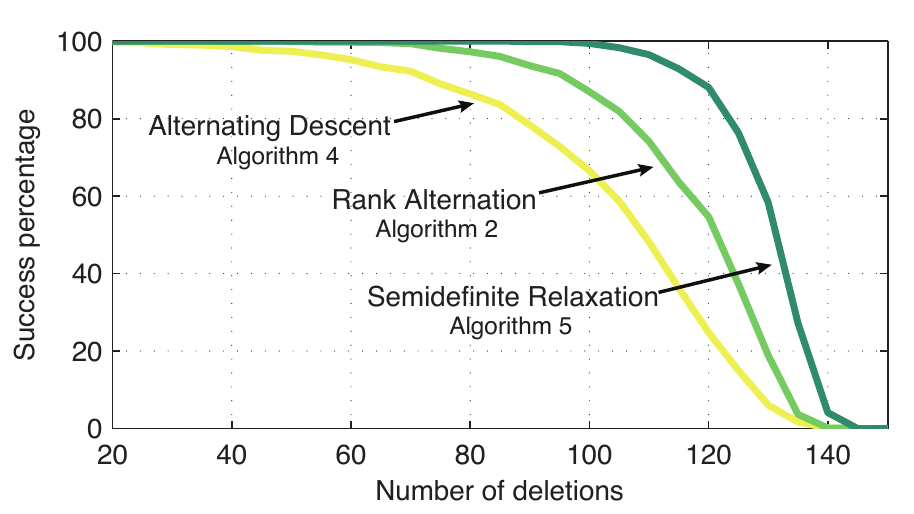}
  \caption{Comparison of different algorithms applied to completing an EDM
  with random deletions. For every number of deletions, we generated 2000
  realizations of 20 points uniformly at random in a unit square. Distances to
  delete were chosen uniformly at random among the resulting $\binom{20}{2} =
  190$ pairs; \rev{20 deletions correspond to $\approx$ 10\% of the number of
  distance pairs and to 5\% of the number of matrix entries; 150 deletions
  correspond to $\approx$ 80\% of the distance pairs and to $\approx$ 38\% of
  the number of matrix entries.} Success was declared if the Frobenius norm of
  the error between the estimated matrix and the true EDM was less than 1\% of
  the Frobenius norm of the true EDM.}
  \label{fig:success_random}
\end{figure}

Denote by $\mR = [\vr_1, \ \ldots, \ \vr_m]$ the unknown microphone locations,
and by $\mS = [\vs_1, \ \ldots, \ \vs_k]$ the unknown source locations. The
distance between the $i$th microphone and $j$th source is
\begin{equation}
	\delta_{ij} = \norm{\vr_i - \vs_j}^2,
\end{equation}
so that in analogy with \eqref{eq:edm_assemble} we have 
\begin{equation}
	\mDelta = \EDM(\mR, \mS) = \diag(\mR^\T \mR) \vone^\T - 2 \mR^\T \mS + \vone \diag(\mS^\T \mS),
\end{equation}
where we overloaded the $\EDM$ operator in a natural way. We use $\mDelta$ to
avoid confusion with the standard Euclidean $\mD$. Consider now two geometric
centering matrices of sizes $m$ and $k$, denoted $\mJ_m$ and $\mJ_k$. Similarly to \eqref{eq:centering_derivation}, we have
\begin{equation}
	\mR \mJ_m = \mR - \vr_c \vone^\T,\ \mS \mJ_k = \mS - \vs_c \vone^\T.
\end{equation}
This means that
\begin{equation}
	\label{eq:quasi_inner_product}
	\mJ_m \mDelta \mJ_k = \wt{\mR}^\T \wt{\mS} \bydef \wt{\mG}
\end{equation}
is a matrix of inner products between vectors $\wt{\vr}_i$ and $\wt{\vs}_j$.
We used tildes to differentiate this from \emph{real} inner products betwen
$\vr_i$ and $\vs_j$, because in \eqref{eq:quasi_inner_product}, the points in
$\wt{\mR}$ and $\wt{\mS}$ are referenced to different coordinate systems. The
centroids $\vr_c$ and $\vs_c$ generally do not coincide. There are different
ways to decompose $\wt{\mG}$ into a product of two full rank matrices, call
them $\mA$ and $\mB$,
\begin{equation}
	\label{eq:quasi_gram_decomposition_AB}
	\wt{\mG} = \mA^\T \mB.
\end{equation}
We could for example use the SVD, $\wt{\mG} = \mU \mSigma \mV^\T$, and set
$\mA^\T = \mU$ and $\mB = \mSigma \mV^\T$. Any two such decompositions are
linked by some invertible transformation $\mT \in \R^{d \times d}$,
\begin{equation}
	\wt{\mG} = \mA^\T \mB = \wt{\mR}^\T \mT^{-1} \mT \wt{\mS}.
\end{equation}
We can now write down the conversion rule from what we can measure to what we
can compute,
\begin{equation}
\begin{aligned}
	&\mR = \mT^\T \mA + \vr_c \vone^\T \\
	&\mS = (\mT^{-1})^\T \mB + \vs_c \vone^\T,
\end{aligned}
\end{equation}
where $\mA$ and $\mB$ can be computed according to
\eqref{eq:quasi_gram_decomposition_AB}. Because we cannot reconstruct the
absolute position of the point set, we can arbitrarily set $\vr_c = 0$, and
$\vs_c = \alpha \ve_1$. Recapitulating, we have that
\begin{equation}
	\label{eq:mdu_equivalent}
	\mDelta = \EDM\left(\mT^\T \! \mA, \ (\mT^{-1})^\T \mB + \alpha \ve_1 \vone^\T\right),
\end{equation}
and the problem is reduced to computing $\mT$ and $\alpha$ so that
\eqref{eq:mdu_equivalent} hold, or in other words, so that the right hand
side be consistent with the data $\mDelta$. We reduced MDU to a relatively
small problem: in 3D, we need to compute only ten scalars. Sch\"onemann
\cite{Schonemann:1970wd} gives an algebraic method to find these parameters,
and mentions the possibility of least squares, while Crocco, Bue and Murino
\cite{Crocco:2012eu} propose a different approach using non-linear least
squares.

This procedure seems quite convoluted. Rather, we see MDU as a special case of
matrix completion, with the structure illustrated in Fig.
\ref{fig:mdu}.

More concretely, represent the microphones and the sources by a set of $n =
k+m$ points, ascribed to the columns of matrix $\mX = [\mR \ \mS]$. Then
$\EDM(\mX)$ has a special structure as seen in Fig. \ref{fig:mdu},
\begin{equation}
	\EDM(\mX) = 
	\begin{bmatrix}
		\EDM(\mR) & \EDM(\mR, \mS) \\
		\EDM(\mS, \mR) & \EDM(\mS)
	\end{bmatrix}.
\end{equation}

We define the mask matrix for MDU as
\begin{equation}
	\mW_{\text{MDU}} \bydef
	\begin{bmatrix}
	    \vzero_{m \times m} & \vone_{m \times k} \\
	    \vone_{k \times m} & \vzero_{k \times k}
	\end{bmatrix}.
\end{equation}
With this matrix, we can simply invoke the SDR in Algorithm
\ref{alg:SDR_matlab}. We could also use Algorithm
\ref{alg:rank_complete_edm}, or Algorithm \ref{alg:alt_coor_desc}. Performance
of different algorithms is compared in Section
\ref{sub:performance_comparison_of_algorithms}.

It is worth mentioning that SNL specific algorithms that exploit the
particular graph induced by limited range communication do not perform well on
MDU. This is because the structure of the missing entries in MDU is in a
certain sense opposite to the one of SNL. %This property is illustrated in Fig. ??.

% subsection box_multidimensional_unfolding (end)

\subsection{Performance Comparison of Algorithms} % (fold)
\label{sub:performance_comparison_of_algorithms}

\begin{figure}[t!]
\centering
   \includegraphics[width=3.5in]{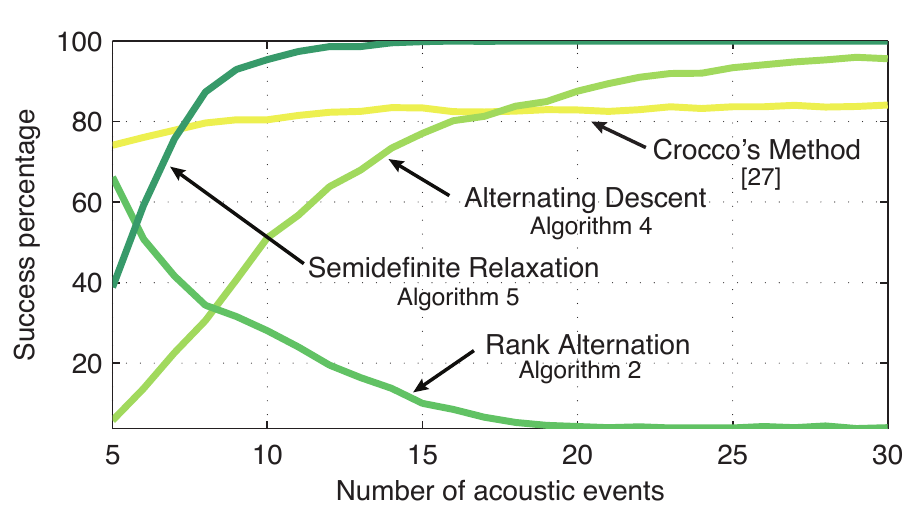}
   \caption{Comparison of different algorithms applied to multidimensional
   unfolding with varying number of acoustic events $k$. For every number of
   acoustic events, we generated 3000 realizations of $m = 20$ microphone
   locations uniformly at random in a unit cube. \rev{Percentage of missing
   matrix entries is given as $(k^2 + m^2) / (k + m)^2$, so that the ticks on
   the abscissa correspond to $[68,56,51,50,51,52]\%$ (non-monotonic in $k$
   with the minimum for $k=m=20$).} Success was declared if the Frobenius norm of
   the error between the estimated matrix and the true EDM was less than 1\%
   of the Frobenius norm of the true EDM.}
   \label{fig:success_mdu}
\end{figure}

We compare the described algorithms in two different EDM completion settings.
In the first experiment (Figs. \ref{fig:success_random} and
\ref{fig:noise_random}), the entries to delete are chosen uniformly at random.
The second experiment (Figs. \ref{fig:success_mdu} and \ref{fig:noise_mdu})
tests performance in MDU, where the non-observed entries are highly
structured. \rev{In Figs. \ref{fig:success_random} and \ref{fig:success_mdu}, we
assume that the observed entries are known exactly, and we plot the success
rate (percentage of accurate EDM reconstructions) against the number of
deletions in the first case, and the number of calibration events in the
second case. Accurate reconstruction is defined in terms of the relative
error. Let $\mD$ be the true, and $\wh{\mD}$ the estimated EDM. The relative
error is then ${\| \wh{\mD} - \mD \|_F}/{\|\mD\|_F}$, and we declare success if
this error is below $1\%$.

To generate Figs. \ref{fig:noise_random} and \ref{fig:noise_mdu} we varied the
amount of random, uniformly distributed jitter added to the distances, and for
each jitter level we plotted the relative error. The exact values of
intermediate curves are less important than the curves for the smallest and
the largest jitter, and the overall shape of the ensemble.

\begin{figure}[t!]
\centering
\includegraphics[width=3.5in]{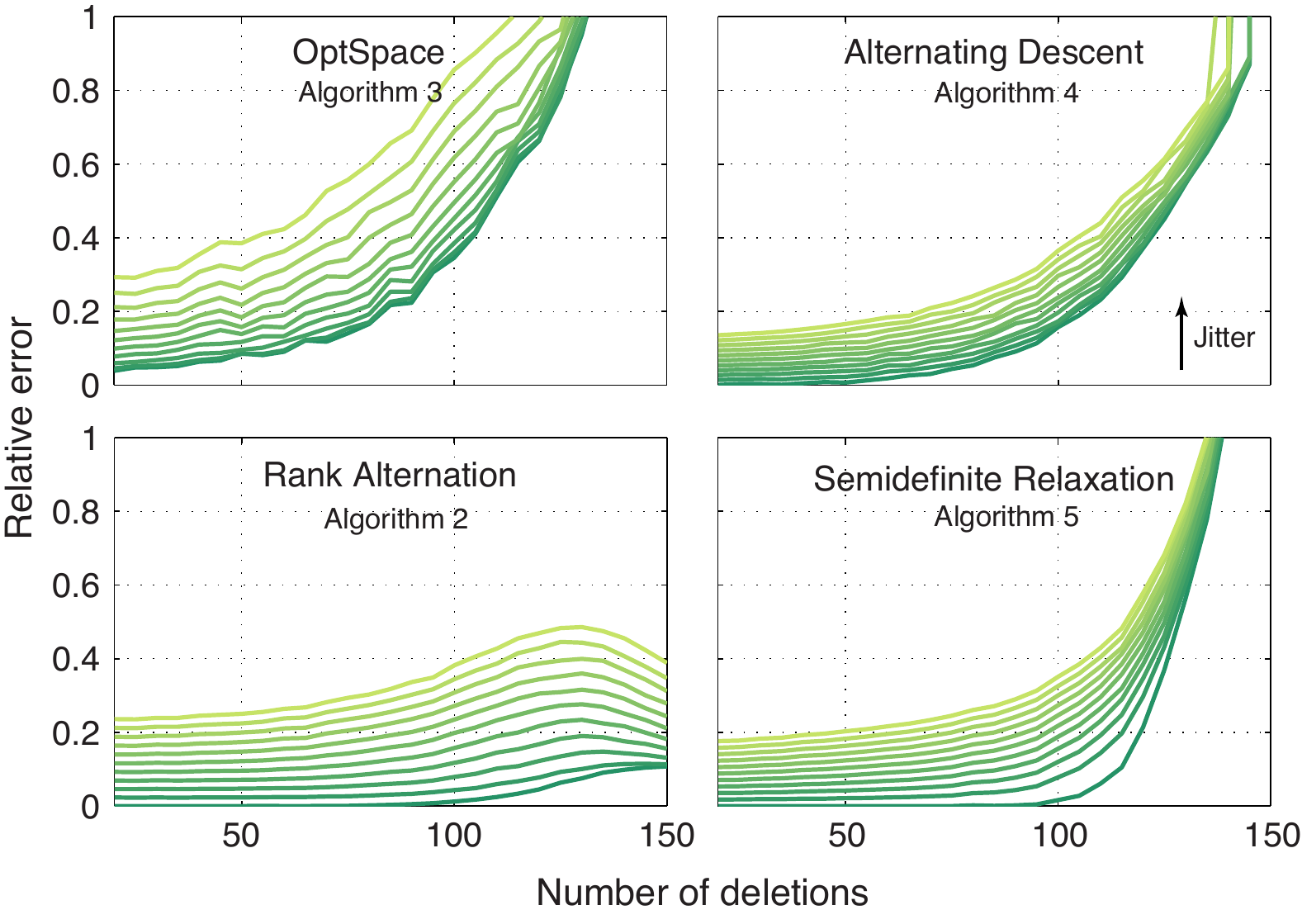}
\caption{\rev{Comparison of different algorithms applied to completing an EDM with
random deletions and noisy distances. For every number of deletions, we
generated 1000 realizations of 20 points uniformly at random in a unit square.
In addition to the number of deletions, we varied the amount of jitter added
to the distances. Jitter was drawn from a centered uniform distribution, with
the level increasing in the direction of the arrow, from $\mathcal{U}[0, 0]$
(no jitter) for the darkest curve at the bottom, to $\mathcal{U}[-0.15, 0.15]$
for the lightest curve at the top, in 11 increments. For every jitter level,
we plotted the mean relative error $\|\wh{\mD}
- \mD \|_F / \|\mD \|_F$ for all algorithms.}}
\label{fig:noise_random}
\end{figure}

A number of observations can be made about the performance of algorithms.
Notably, OptSpace (Algorithm \ref{alg:optspace}) does not perform well
for randomly deleted entries when $n = 20$; it was designed for larger
matrices. For this matrix size, the mean relative reconstruction error
achieved by OptSpace is the worst of all algorithms (Fig.
\ref{fig:noise_random}). In fact, the relative error in the noiseless case was
rarely below the success threshold (set to $1
\%$) so we omitted the corresponding near-zero curve from Fig.
\ref{fig:success_random}. Furthermore, OptSpace assumes that the pattern of
missing entries is random; in the case of a blocked deterministic structure
associated with MDU, it never yields a satisfactory completion.

On the other hand, when the unobserved entries are randomly scattered in the
matrix, and the matrix is large---in the ultrasonic calibration example the
number of sensors $n$ was $200$ or more---OptSpace is a very fast and
attractive algorithm. To fully exploit OptSpace, $n$ should be even larger, in
the thousands or tens of thousands.

SDR (Algorithm \ref{alg:SDR_matlab}) performs well in all scenarios. For both
the random deletions and the MDU, it has the highest success rate, and it
behaves well with respect to noise. Alternating coordinate descent (Algorithm
\ref{alg:alt_coor_desc}) performs slightly better in noise for small number of
deletions and large number of calibration events, but Figs.
\ref{fig:success_random} and \ref{fig:success_mdu} indicate that for certain
realizations of the point set it gives large errors. If the worst-case
performance is critical, SDR is a better choice. We note that in the
experiments involving the SDR, we have set the multiplier $\lambda$ in
\ref{eq:sdp_trace_maximization} to the square root of the number of missing
entries. This choice was empirically found to perform well.

The main drawback of SDR is speed; it is the slowest among the tested
algorithms. To solve the semidefinite program we used CVX \cite{cvx,gb08}, a
Matlab interface to various interior point methods. For larger matrices
(\emph{e.g.}, $n = 1000$), CVX runs out of memory on a desktop computer, and
essentially never finishes. Matlab implementations of alternating coordinate
descent, rank alternation (Algorithm \ref{alg:rank_complete_edm}), and
OptSpace are all much faster.

The microphone calibration algorithm by Crocco \cite{Crocco:2012eu} performs
equally well for any number of acoustic events. This may be explained by the
fact that it always reduces the problem to ten unknowns. It is an attractive
choice for practical calibration problems with a smaller number of calibration
events. Algorithm's success rate can be further improved if one is prepared to
run it for many random initializations of the non-linear
optimization step.

\begin{figure}[t!]
\centering
\includegraphics[width=3.5in]{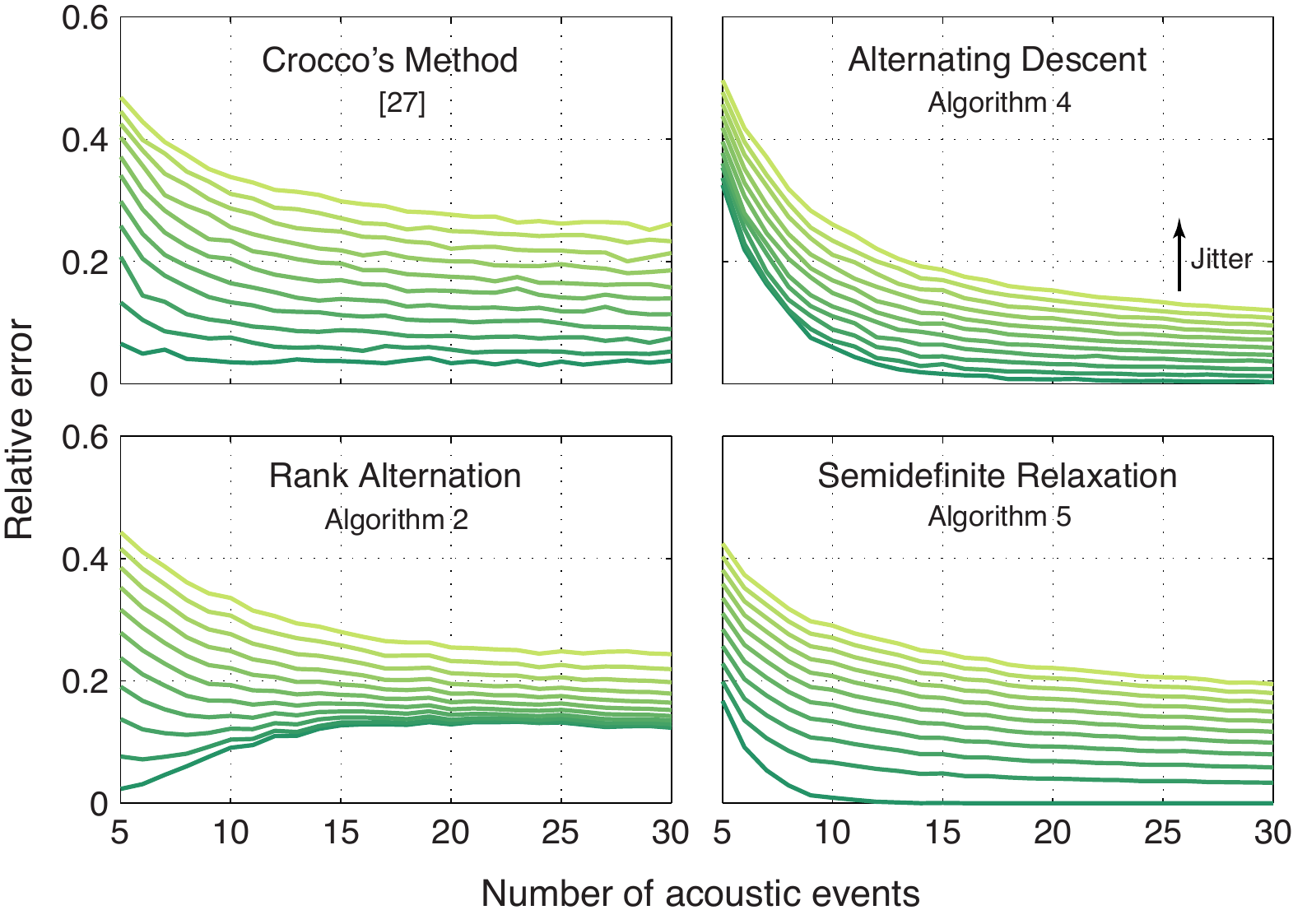}
\caption{\rev{Comparison of different algorithms applied to multidimensional
unfolding with varying number of acoustic events $k$ and noisy distances. For
every number of acoustic events, we generated 1000 realizations of $m = 20$
microphone locations uniformly at random in a unit cube. In addition to the
number of acoustic events, we varied the amount of random jitter added to the
distances, with the same parameters as in Fig.~\ref{fig:noise_random}. For
every jitter level, we plotted the mean relative error $\|\wh{\mD} - \mD \|_F /
\|\mD \|_F$ for all algorithms.}}
\label{fig:noise_mdu}
\end{figure}

Interesting behavior can be observed for the rank alternation in MDU. Figs.
\ref{fig:success_mdu} and \ref{fig:noise_mdu} both show that at low noise
levels, the performance of the rank alternation becomes worse with the number
of acoustic events. At first glance, this may seem counterintuitive, as more
acoustic events means more information; one could simply ignore some of them,
and perform at least equally well as with fewer events. But this reasoning
presumes that the method is \emph{aware} of the geometrical meaning of the
matrix entries; on the contrary, rank alternation is using only rank.
Therefore, even if the percentage of the observed matrix entries grows until a
certain point, the size of the structured blocks of unknown entries grows as
well (and the percentage of known entries in columns/rows corresponding to
acoustic events decreases). This makes it harder for a method that does not
use geometric relationships to complete the matrix. A loose comparison can be
made to image inpainting: If the pixels are missing randomly, many methods
will do a good job; but if a large patch is missing, we cannot do much without
additional structure (in our case geometry), no matter how large the rest of
the image is.

To summarize, for smaller matrices the SDR seems to be the best overall
choice. For large matrices the SDR becomes too slow and one should turn to
alternating coordinate descent, rank alternation or OptSpace. Rank alternation
is the simplest algorithm, but alternating coordinate descent performs better.
For very large matrices ($n$ on the order of thousands or tens of thousands),
OptSpace becomes the most attractive solution.} We note that we deliberately
refrained from making detailed running time comparisons, due to the diverse
implementations of the algorithms.

% subsection performance_comparison_of_algorithms (end)

\subsection{Summary} % (fold)

In this section we discussed:
\begin{itemize}
	\item The problem statement for EDM completion and denoising, and how to
	easily exploit the rank property (Algorithm
	\ref{alg:rank_complete_edm}),
	\item Standard objective functions in MDS: raw stress and s-stress, and
	 a simple algorithm to minimize s-stress (Algorithm
	\ref{alg:alt_coor_desc}),
	\item Different semidefinite relaxations that exploit the connection
	between EDMs and PSD matrices,
	\item Multidimensional unfolding, and how to solve it efficiently using
	EDM completion,
  \item Performance of the introduced algorithms in two very different
  scenarios: EDM completion with randomly unobserved entries, and EDM
  completion with deterministic block structure of unobserved entries (MDU).
\end{itemize}

% subsection summary (end)

% section edms_as_a_practical_tool (end)

\section{Unlabeled Distances} % (fold)
\label{sec:unlabeled_distances}

\begin{figure}[t]
\centering
\includegraphics[width=3.5in]{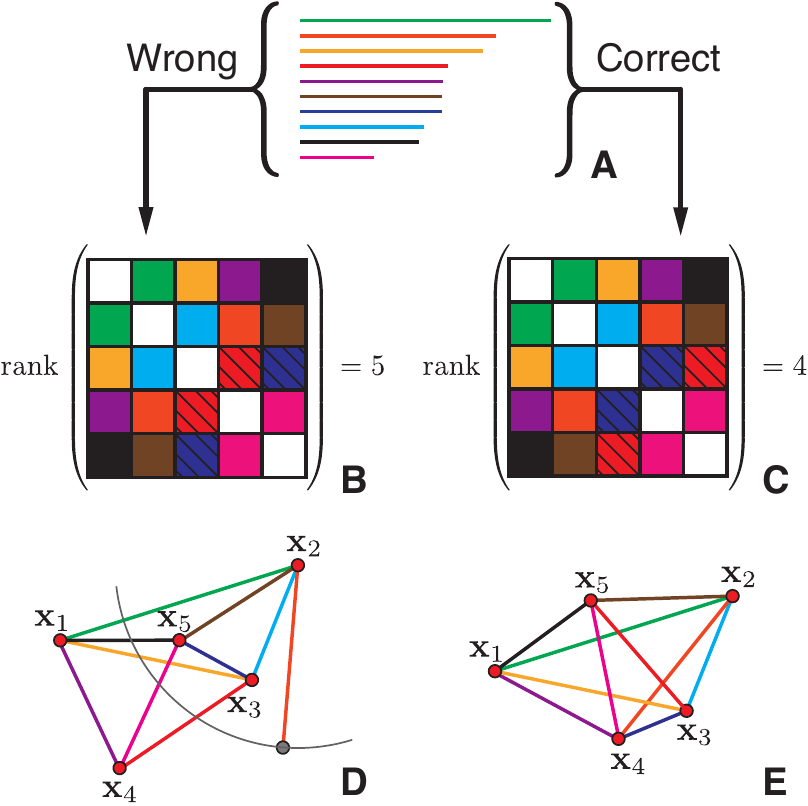}
\caption{Illustration of the uniqueness of EDMs for unlabeled distances. A set
of unlabeled distance (A) is distributed in two different ways in a tentative
EDM with embedding dimension 2 (B and C). The correct assignment yields the
matrix with the expected rank (C), and the point set is easily realized in the
plane (E). On the contrary, swapping just two distances (hatched squares in
(B) and (C)) makes it impossible to realize the point set in the plane (D).
Triangles that do not coincide with the swapped edges can still be placed, but
in the end we are left with a hanging orange stick that cannot attach itself
to any of the five nodes.}
\label{fig:unlabeled_distances}
\end{figure}

In certain applications we can measure the distances between the points, but
we do not know the correct labeling. That is, we know all the entries of an
EDM, but we do not know how to arrange them in the matrix. As illustrated in
Fig. \ref{fig:unlabeled_distances}\textsl{A}, we can imagine having a set of
sticks of various lengths. The task is to work out the correct way to connect
the ends of different sticks so that no stick is left hanging open-ended.

In this section we exploit the fact that in many cases, distance labeling is
not essential. For most point configurations, there is no other set of points
that can generate the corresponding set of distances, up to a rigid
transformation.

Localization from unlabeled distances is relevant in various calibration
scenarios where we cannot tell apart distance measurements belonging to
different points in space. This can occur when we measure times of arrivals of
echoes, which correspond to distances between the microphones and the image
sources (see Fig. \ref{fig:room_hearing_is})
\cite{Dokmanic:2014tc, Dokmanic:2013dz}. Somewhat surprisingly, the same
problem of unlabeled distances appears in sparse phase retrieval; to see how,
take a look at the ``\textbf{Phase Retrieval}'' box.

No efficient algorithm currently exists for localization from unlabeled
distances in the general case of noisy distances. \rev{We should mention,
however, a recent polynomial-time algorithm (albeit of a high degree) by
Gujarathi and \emph{et al.} \cite{Gujarathi:2014cz}, that can reconstruct
relatively large point sets from unordered, noiseless distance data.}

At any rate, the number of assignments to test is sometimes sufficiently small
so that an exhaustive search does not present a problem. We can then use EDMs
to find the best labeling. The key to the unknown permutation problem is the
following fact.

\begin{thm}
	\label{thm:unlabeled}
  Draw $\vx_1, \vx_2, \cdots, \vx_n \in \R^d$ independently from some
  absolutely continuous probability distribution (\emph{e.g.} uniformly at
  random) on $\Omega \subseteq \R^d$. Then with probability 1, the obtained
  point configuration is the unique \rev{(up to a rigid transformation)} point
  configuration in $\Omega$ that generates the set of distances
  $\set{\norm{\vx_i -	\vx_j}, 1 \leq i < j \leq n}$.
\end{thm}

This fact is a simple consequence of a result by Boutin and Kemper
\cite{Boutin:2004gb} who give a characterization of point sets reconstructible
from unlabeled distances.

Figs. \ref{fig:unlabeled_distances}\textsl{B} and
\ref{fig:unlabeled_distances}\textsl{C} show two possible arrangements of the
set of distances in a tentative EDM; the only difference is that the two
hatched entries are swapped. But this simple swap is not harmless: there is no
way to attach the last stick in Fig. \ref{fig:unlabeled_distances}\textsl{D},
while keeping the remaining triangles consistent. We could do it in a higher
embedding dimension, but we insist on realizing it in the plane.

What Theorem \ref{thm:unlabeled} does not tell us is how to identify the
correct labeling. But we know that for most sets of distances, only one
(correct!) permutation can be realized in the given embedding dimension. Of
course, if all the labelings are unknown and we have no good heuristics to
trim the solution space, finding the correct labeling is difficult, as noted
in \cite{Gujarathi:2014cz}. Yet there are interesting situations where this
search is feasible because we can augment the EDM point by point. We describe
one such situation in the next subsection.

%----------------------------------------------------------------------------
% PR box
%----------------------------------------------------------------------------
\begin{figure*}[t]
\fontfamily{lmss}\selectfont
\begin{spmagbox}
\textbf{EDM Perspective on Sparse Phase Retrieval (The Unexpected Distance Structure)}
\\
\label{sub:box_phase_retrieval_by_edms}
\fontsize{9pt}{10pt}\selectfont
\begin{multicols}{2}

In many cases, it is easier to measure a signal in the Fourier domain.
Unfortunately, it is common in these scenarios that we can only reliably
measure the magnitude of the Fourier transform (FT). We would like to recover
the signal of interest from just the magnitude of its FT, hence the name
\emph{phase retrieval}. X-ray crystallography
\cite{Millane:1990pt} and speckle imaging in astronomy
\cite{Beavers:1989tj} are classic examples of phase retrieval problems. In
both of these applications the signal is spatially sparse. We can model it as

\begin{align} 
	f(\vx)=\sum_{i=1}^n c_i \delta(\vx-\vx_i),
\end{align} 

where $c_i$ are the amplitudes and $\vx_i$  the locations of the $n$ Dirac
deltas in the signal. In what follows, we discuss the problem on
1-dimensional domains, that is $\vx\in\R$, knowing that a multidimensional
phase retrieval problem can be solved by solving many 1-dimensional problems
\cite{Ranieri:2013tx}.

Note that measuring the magnitude of the FT of $f(\vx)$ is equivalent to
measuring its autocorrelation function (ACF). For a sparse $f(\vx)$, the ACF
is also sparse and given as
\begin{align}
    a(\vx)=\sum_{i=1}^n\sum_{j=1}^nc_ic_j\delta(\vx-(\vx_i-\vx_j)),
\end{align}
where we note the presence of differences between the locations $\vx_i$ in the
support of the ACF. As $a(\vx)$ is symmetric, we do not know the order of
$\vx_i$, and so we can only know these differences up to a sign, which is
equivalent to knowing the distances $\norm{\vx_i - \vx_j}$.

For the following reasons, we focus on the recovery of the support of the
signal $f(\vx)$ from the support of the ACF $a(\vx)$: i) in certain
applications, the amplitudes $c_i$ may be all equal, thus limiting their role
in the reconstruction; ii) knowing the support of $f(\vx)$ and its ACF
is sufficient to exactly recover the signal $f(\vx)$
\cite{Ranieri:2013tx}. 

The recovery of the support of $f(\vx)$ from the one of $a(\vx)$ corresponds
to the localization of a set of $n$ points from their \emph{unlabeled}
distances: we have access to all the pairwise distances but we do not know
which pair of points corresponds to any given distance. This can be recognized
as an instance of the \emph{turnpike problem}, whose computational complexity
is believed not to be NP-hard but for which no polynomial time algorithm is
known \cite{Lemke:2002um}.

From an EDM perspective, we can design a reconstruction algorithm recovering
the support of the signal $f(\vx)$ by labeling the distances obtained from the
ACF such that the resulting EDM has rank smaller or equal than 3. This can be
regarded as unidimensional scaling with unlabeled distances, and the algorithm
to solve it is similar to echo sorting (Algorithm \ref{alg:echo_sorting}).

\vspace{5mm}

\includegraphics[width=3.1in]{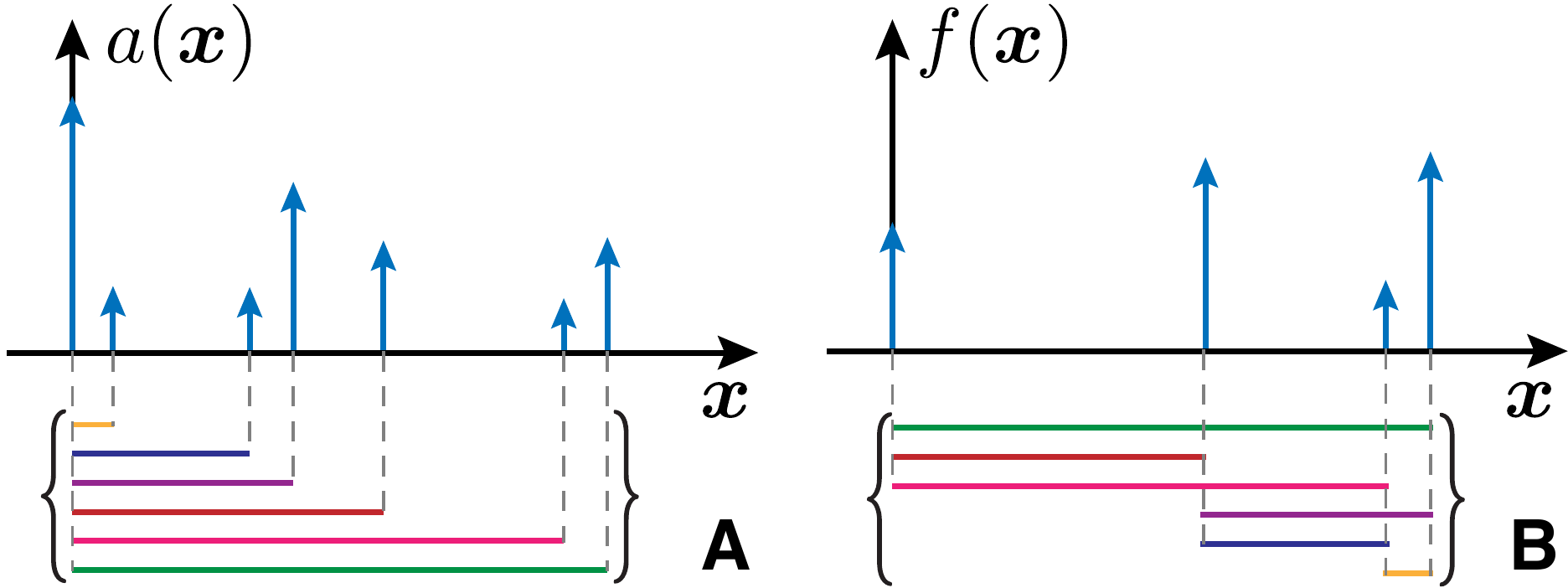}

\caption{\fontfamily{lmss}\selectfont A graphical representation of the phase retrieval problem for
1-dimensional sparse signals. (A) We measure the ACF of the signal and we
recover a set of distances (\emph{sticks} in Fig.
\ref{fig:unlabeled_distances}) from its support. (B) These are the unlabeled
distances between all the pairs of Dirac deltas in the signal $f(\vx)$. We
exactly recover the support of the signal if we correctly label the distances.}

\end{multicols}
\end{spmagbox}
\end{figure*}

\subsection{Hearing the Shape of a Room \protect{\cite{Dokmanic:2013dz}}} % (fold)
\label{sub:hearing_the_shape_of_a_room}

\begin{figure}[t]
\centering
\includegraphics[width=3.4in]{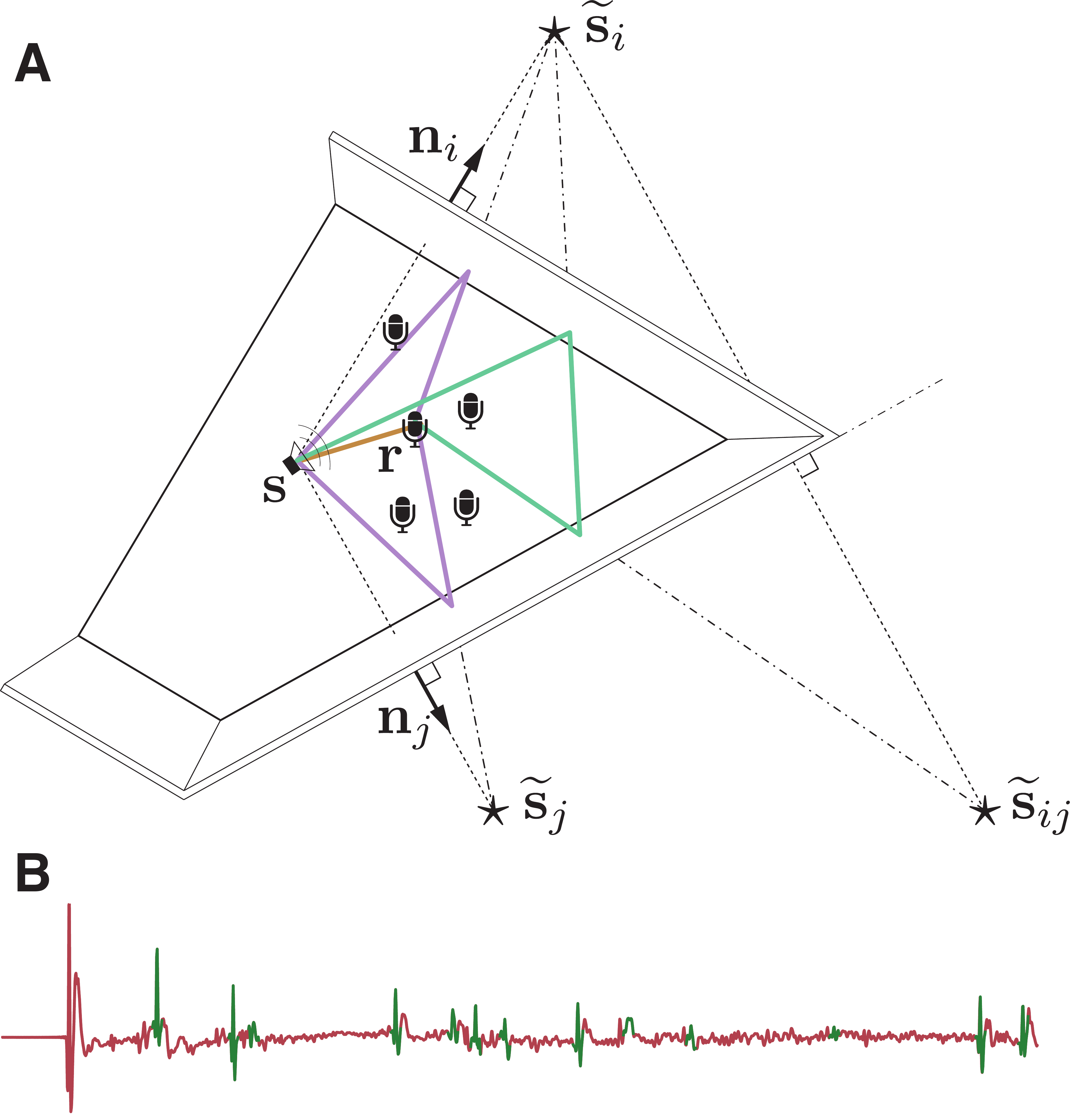}
\caption{(A) Illustration of the image source model for first- and second-order
echoes. Vector $\vn_i$ is the outward-pointing unit normal associated with the
$i$th wall. Stars denote the image sources, and $\wt{\vs}_{ij}$ is the image
source corresponding to the second-order echo. Sound rays corresponding to
first reflections are shown in blue, and the ray corresponding to the
second-order reflection is shown in red. (B) Early part of a typical recorded
room impulse response.}
\label{fig:room_hearing_is}
\end{figure}

An important application of EDMs with unlabeled distances is the
reconstruction of the room shape from echoes. An acoustic setup is shown in
Fig. \ref{fig:room_hearing_is}\textsl{A}, but one could also use radio
signals. Microphones pick up the convolution of the sound emitted by the
loudspeaker with the room impulse response (RIR), which can be estimated by
knowing the emitted sound. An example RIR recorded by one of the microphones
is illustrated in Fig.
\ref{fig:room_hearing_is}\textsl{B}, with peaks highlighted in green.
Some of these peaks are first-order echoes coming from different walls, and
some are higher-order echoes or just noise.

Echoes are linked to the room geometry by the image source model
\cite{Allen:1979ua}. According to this model, we can replace echoes by image
sources (IS)---mirror images of the true sources across the corresponding walls.
Position of the image source of $\vs$ corresponding to wall $i$ is computed as

\begin{equation}
\label{eq:image-sources}
\wt{\vs}_i = \vs + 2 \inprod{\vp_i - \vs, \vn_i} \vn_i,
\end{equation}
where $\vp_i$ is any point on the $i$th wall, and $\vn_i$ is the unit normal
vector associated with the $i$th wall, see Fig.
\ref{fig:room_hearing_is}\textsl{A}.
A convex room with planar walls is completely determined by the locations of
first-order ISs \cite{Dokmanic:2013dz}, so by reconstructing their locations,
we actually reconstruct the room's geometry.

We assume that the loudspeaker and the microphones are synchronized so that
the times at which the echoes arrive directly correspond to distances. The
challenge is that the distances---the green peaks in
Fig.~\ref{fig:room_hearing_is}\textsl{B}---are unlabeled: it might happen that
the $k$th peak in the RIR from microphone 1 and the $k$th peak in the RIR from
microphone 2 come from different walls, especially for larger microphone
arrays. Thus, we have to address the problem of echo sorting, in order to
group peaks corresponding to the same image source in RIRs from different
microphones.

Assuming that we know the pairwise distances between the microphones $\mR =
[\vr_1, \ldots, \vr_m]$, we can create an EDM corresponding to the microphone
array. Because echoes correspond to image sources, and image sources are just
points in space, we attempt to grow that EDM by adding one point---an image
source---at a time. To do that, we pick one echo from every microphone's
impulse response, augment the EDM based on echo arrival times, and check how
far the augmented matrix is from an EDM with embedding dimension three, as we
work in 3D space. The distance from an EDM is measured with s-stress cost
function. It was shown in \cite{Dokmanic:2013dz} that a variant of Theorem
\ref{thm:unlabeled} applies to image sources when microphones are thrown at
random. Therefore, if the augmented matrix satisfies the EDM properties,
almost surely we have found a good image source. With probability 1, no
other combination of points could have generated the used distances.

The main reason for using EDMs and s-stress instead of, for instance, the rank
property, is that we get robust algorithms. Echo arrival times are corrupted
with various errors, and relying on the rank is too brittle. It was verified
experimentally \cite{Dokmanic:2013dz} that EDMs and s-stress yield a very
robust filter for the correct combinations of echoes.

Thus we may try all feasible combinations of echoes, and expect to get exactly
one ``good'' combination for every image source that is ``visible'' in the
impulse responses. In this case, as we are only adding a single point, the
search space is small enough to be rapidly traversed exhaustively. Geometric
considerations allow for a further trimming of the search space: \rev{because we
know the diameter of the microphone array, we know that an echo from a
particular wall must arrive at all the microphones within a temporal window
corresponding to the array's diameter.}

The procedure is as follows: collect all echo arrival times received by the
$i$th microphone in the set $T_i$, and fix $t_1 \in T_1$ corresponding to a
particular image source. Then Algorithm \ref{alg:echo_sorting} finds echoes in
other microphones' RIRs that correspond to this same image source. Once we
group all the peaks corresponding to one image source, we can determine its
location by multilateration (\emph{e.g.} by running the classical MDS), and
then repeat the process for other echoes in $T_1$.

\rev{To get a ballpark idea of the number of combinations to test, suppose that we
detect 20 echoes per microphone\footnote{\rev{We do not need to look beyond
early echoes corresponding to at most three bounces. This is convenient, as
echoes of higher orders are challenging or impossible to isolate.}}, and that the
diameter of the five-microphone array is $1.5$ m. Thus for every peak time
$t_1 \in T_1$ we have to look for peaks in the remaining four microphones that arrived
within a window around $t_1$ of length $2 \times
\frac{1 \text{~m}}{343 \text{~m/s}}$, where $343$ m/s is the
speed of sound. This is approximately $6$ ms, and in a typical room we can
expect about five early echoes within a window of that duration. Thus we have
to compute the s-stress for $20 \times 5^4 = 12500$ matrices of size $6 \times
6$, which can be done in a matter of seconds on a desktop computer. In fact,
once we assign an echo to an image source, we can exclude it from further
testing, so the number of combinations can be further reduced.}

\begin{algorithm}[t]
\caption{Echo Sorting \cite{Dokmanic:2013dz}}
\label{alg:echo_sorting}
\begin{algorithmic}[1]
\Function{EchoSort}{$\mR, t_1, \ldots, T_m$}
	\State $\mD \gets \EDM(\mR)$
	\State $s_\mathrm{best} \gets +\mathrm{Inf}$
	\ForAll{$\vt = [t_2, \ldots, t_m], \ \text{such that}~t_i \in T_i$}
		\State $\vd \gets c \cdot [t_1, \ \vt^\T]^\T$ \Comment{$c$ is the sound speed}
		\State $\mD_\text{aug} \gets \begin{bmatrix} \mD & \vd \\ \vd^\T & 0 \end{bmatrix}$
		\If{$\mathrm{s\!-\!stress}(\mD_\mathrm{aug}) < s_\mathrm{best}$}
			\State $s_\mathrm{best} \gets \mathrm{s\!-\!stress}(\mD_\mathrm{aug})$
			\State $\vd_\mathrm{best} \gets \vd$
		\EndIf
	\EndFor
	\State \Return $\vd_\mathrm{best}$
\EndFunction
\end{algorithmic}
\end{algorithm}
Algorithm \ref{alg:echo_sorting} was used to reconstruct rooms with centimeter
precision \cite{Dokmanic:2013dz} with one loudspeaker and an array of five
microphones. The same algorithm also enables a dual application: indoor
localization of an acoustic source using only one microphone---a feat not
possible if we are not in a room \cite{Parhizkar:2014kn}.

% subsection hearing_the_shape_of_a_room (end)

\subsection{Summary} % (fold)

To summarize this section:
\begin{itemize}
	\item We explained that for most point sets, the distances they generate
	are unique; there are no other point sets generating the same distances,
  \item In room reconstruction from echoes, we need to identify the correct
  labeling of the distances to image sources. EDMs act as a robust filter for
  echoes coming from the same image source,
	\item Sparse phase retrieval can be cast as a distance problem, too. The
	support of the ACF gives us distances between the deltas in the original
	signal. Echo sorting can be adapted to solve the problem from the EDM
	perspective.
\end{itemize}

% subsection summary (end)

\begin{table}[tb]
\centering
\caption{Applications of EDMs with different twists}
\begin{tabular}{@{}r | p{1.4cm} p{1.4cm} p{1.4cm} }
\toprule[1.2pt]
Application & missing \hspace{1mm}distances & noisy distances & unlabeled distances \\
\midrule
%\addlinespace[8pt]
Wireless sensor networks  & \ding{52} & \ding{52} & $\times$ \\
Molecular conformation & \ding{52} & \ding{52} & $\times$ \\
Hearing the shape of a room & $\times$ & \ding{52} & \ding{52}  \\
Indoor localization & $\times$ & \ding{52} & \ding{52} \\
Calibration & \ding{52} & \ding{52} & $\times$ \\
Sparse phase retrieval & $\times$ & \ding{52} & \ding{52} \\
  \bottomrule[1.2pt]
\end{tabular}

\label{tab:EDMuncer}
\end{table}

\section{Ideas for Future Research} % (fold)
\label{sec:ideas_for_future_research}

Even problems that at first glance seem to have little to do with EDMs,
sometimes reveal a distance structure when you look closely. A good example is
sparse phase retrieval.

The purpose of this paper is to convince the reader that Euclidean distance
matrices are powerful objects with a multitude of applications (Table
\ref{tab:EDMuncer} lists various flavors), and that they should belong to any
practitioner's toolbox. We have an impression that the power of EDMs and the
associated algorithms has not been sufficiently recognized in the signal
processing community, and our goal is to provide a good starting reference. To
this end, and perhaps to inspire new research directions, we list several
EDM-related problems that we are curious about and believe are important.

\paragraph{Distance matrices on manifolds} If the points lie on a particular
manifold, what can be said about their distance matrix? We know that if the
points are on a circle, the EDM has rank three instead of four, and this
generalizes to hyperspheres \cite{gower1}. But what about more general
manifolds? Are there invertible transforms of the data or of the Gram matrix
that yield EDMs with a lower rank than the embedding dimension suggests? What
about different distances, \emph{e.g.} the geodesic distance on the manifold? Answers
to these questions have immediate applications in machine learning, where the
data can be approximately assumed to be on a smooth surface
\cite{Tenenbaum2000}.

\paragraph{Projections of EDMs on lower dimensional subspaces} What happens
to an EDM when we project its generating points to a lower dimensional space?
What is the minimum number of projections that we need to be able to
reconstruct the original point set? Answers to these questions have
significant impact on imaging applications such as X-ray crystallography and
seismic imaging. What happens when we only have partial distance observations
in various subspaces? What are other useful low-dimensional structures on
which we can observe the high-dimensional distance data?

\paragraph{Efficient algorithms for distance labeling} Without
application-specific heuristics to trim down the search space, identifying
correct labeling of the distances quickly becomes an arduous task. Can we
identify scenarios for which there are efficient labeling algorithms? What
happens when we do not have the labeling, but we also do not have the complete
collection of \emph{sticks}? What can we say about uniqueness of incomplete
unlabeled distance sets? \rev{Some of the questions have been answered by
Gujarathi \cite{Gujarathi:2014cz}, but many remain. The quest is on for faster
algorithms, as well as algorithms that can handle noisy distances.

In particular, if the noise distribution on the unlabeled distances is known,
what can we say about the distribution of the reconstructed point set (taking
in some sense the \emph{best} reconstruction over all labelings)? Is it
compact, or we can \emph{jump} to totally wrong assignments with positive
probability?}

\paragraph{Analytical local minimum of s-stress} Everyone agrees that there
are many, but to the best of our knowledge, no analytical minimum of s-stress
has yet been found.

% section ideas_for_future_research (end)

\section{Conclusion}

At the end of this tutorial, we hope that we succeeded in showing how
universally useful EDMs are, and that we inspired readers coming across this
material for the first time to dig deeper. Distance measurements are so common
that a simple, yet sophisticated tool like EDMs deserves attention. A good
example is the semidefinite relaxation: even though it is generic, it is the
best performing algorithm for the specific problem of ad-hoc microphone array
localization. Continuing research on this topic will bring new revolutions,
like it did in the 80s in crystallography. Perhaps the next one will be fueled
by solving the labeling problem.

\section*{Acknowledgments}

We would like to thank Dr. Farid M. Naini: without his help, the numerical
simulations for this paper would have taken forever. We would also like to
thank the anonymous reviewers for their numerous insightful suggestions that
have improved the revised manuscript.

% \bibliographystyle{IEEEtran}
% \bibliography{SPMag_PDV_Refs,ivanbib,rezabib,Juri_Refs}

\begin{thebibliography}{10}
\providecommand{\url}[1]{#1}
\csname url@samestyle\endcsname
\providecommand{\newblock}{\relax}
\providecommand{\bibinfo}[2]{#2}
\providecommand{\BIBentrySTDinterwordspacing}{\spaceskip=0pt\relax}
\providecommand{\BIBentryALTinterwordstretchfactor}{4}
\providecommand{\BIBentryALTinterwordspacing}{\spaceskip=\fontdimen2\font plus
\BIBentryALTinterwordstretchfactor\fontdimen3\font minus
  \fontdimen4\font\relax}
\providecommand{\BIBforeignlanguage}[2]{{%
\expandafter\ifx\csname l@#1\endcsname\relax
\typeout{** WARNING: IEEEtran.bst: No hyphenation pattern has been}%
\typeout{** loaded for the language `#1'. Using the pattern for}%
\typeout{** the default language instead.}%
\else
\language=\csname l@#1\endcsname
\fi
#2}}
\providecommand{\BIBdecl}{\relax}
\BIBdecl

\bibitem{Patwari:2005kc}
N.~Patwari, J.~N. Ash, S.~Kyperountas, A.~O. Hero, R.~L. Moses, and N.~S.
  Correal, ``{Locating the Nodes: Cooperative Localization in Wireless Sensor
  Networks},'' \emph{IEEE Signal Process. Mag.}, vol.~22, no.~4, pp. 54--69,
  Jul. 2005.

\bibitem{Alfakih1999}
A.~Y. Alfakih, A.~Khandani, and H.~Wolkowicz, ``Solving {E}uclidean Distance
  Matrix Completion Problems via Semidefinite Programming,''
  \emph{Comput. Optim. Appl.}, vol.~12, no. 1-3, pp.
  13--30, Jan. 1999.

\bibitem{dohetry2001}
L.~Doherty, K.~Pister, and L.~El~Ghaoui, ``Convex Position Estimation in
  Wireless Sensor Networks,'' in \emph{Proc. IEEE INFOCOM}, vol.~3, 2001, pp.
  1655--1663.

\bibitem{Biswas2004}
P.~Biswas and Y.~Ye, ``Semidefinite Programming For Ad Hoc Wireless Sensor
  Network Localization,'' in \emph{Proc. ACM/IEEE IPSN}, 2004, pp. 46--54.

% \bibitem{Williamson1985}
% M.~P. Williamson, T.~F. Havel, and K.~W\"{u}thrich, ``Solution Conformation of
%   Proteinase Inhibitor {IIA} from Bull Seminal Plasma by 1h Nuclear Magnetic
%   Resonance and Distance Geometry,'' \emph{J. Mol. Biol.}, vol.
%   182, no.~2, pp. 295--315, 1985.

\bibitem{Havel1985281}
T.~F. Havel and K.~W\"{u}thrich, ``An Evaluation of the Combined Use of Nuclear
  Magnetic Resonance and Distance Geometry for the Determination of Protein
  Conformations in Solution,'' \emph{J. Mol. Biol.}, vol. 182,
  no.~2, pp. 281--294, 1985.

% \bibitem{Antonacci:2012hc}
% F.~Antonacci, J.~Filos, M.~R.~P. Thomas, E.~A.~P. Habets, A.~Sarti, P.~A.
%   Naylor, and S.~Tubaro, ``{Inference of Room Geometry From Acoustic Impulse
%   Responses},'' \emph{IEEE Trans. Acoust., Speech, Signal Process.}, vol.~20,
%   no.~10, pp. 2683--2695, 2012.

% \bibitem{Ribeiro:2012ge}
% F.~Ribeiro, D.~A. Florencio, D.~E. Ba, and C.~Zhang, ``{Geometrically
%   Constrained Room Modeling With Compact Microphone Arrays},'' \emph{IEEE
%   Trans. Acoust., Speech, Signal Process.}, vol.~20, no.~5, pp. 1449--1460,
%   2012.

\bibitem{Dokmanic:2013dz}
I.~Dokmani{\'c}, R.~Parhizkar, A.~Walther, Y.~M. Lu, and M.~Vetterli,
  ``{Acoustic Echoes Reveal Room Shape},'' \emph{Proc. Natl. Acad. Sci.}, vol.
  110, no.~30, Jun. 2013.

% \bibitem{Dokmanic:2011vc}
% I.~Dokmani{\'c}, Y.~M. Lu, and M.~Vetterli, ``{Can One Hear the Shape of a Room:
%   The 2-D Polygonal Case},'' in \emph{Proc. IEEE ICASSP}, Prague, 2011, pp. 321--324.

\bibitem{Ranieri:2013tx}
J.~Ranieri, A.~Chebira, Y.~M. Lu, and M.~Vetterli, ``{Phase Retrieval for
  Sparse Signals: Uniqueness Conditions},'' \emph{submitted to IEEE Trans. Inf.
  Theory}, Jul. 2013.

\bibitem{torgerson1952}
W.~S. Torgerson, ``Multidimensional Scaling: I. Theory and Method,''
  \emph{Psychometrika}, vol.~17, pp. 401--419, 1952.

\bibitem{Weinberger2004}
K.~Q. Weinberger and L.~K. Saul, ``Unsupervised Learning of Image Manifolds by
  Semidefinite Programming,'' in \emph{Proc. IEEE CVPR}, 2004.

\bibitem{Liberti:2012ut}
L.~Liberti, C.~Lavor, N.~Maculan, and A.~Mucherino, ``{Euclidean Distance
  Geometry and Applications},'' \emph{SIAM Rev.}, vol.~56, no.~1, pp. 3--69,
  2014.

\bibitem{Menger:1928bc}
K.~Menger, ``{Untersuchungen {\"U}ber Allgemeine Metrik},'' \emph{Math. Ann.},
  vol. 100, no.~1, pp. 75--163, Dec. 1928.

% \bibitem{Menger:1931jn}
% ------, ``{New Foundation of Euclidean Geometry},'' \emph{Amer. J. Math.},
% vol.~53, no.~4, p. 721, Oct. 1931.

\bibitem{Schoenberg:1935dk}
I.~J. Schoenberg, ``{Remarks to Maurice Frechet's Article ``Sur La D\'{e}finition
  Axiomatique D'Une Classe D'Espace Distanc\'{e}s Vectoriellement Applicable Sur
  L'Espace De Hilbert},'' \emph{Ann. Math.}, vol.~36, no.~3, p. 724, Jul. 1935.


\bibitem{Blumenthal:1953ie}
L.~M. Blumenthal, \emph{{Theory and Applications of Distance Geometry}}.\hskip
  1em plus 0.5em minus 0.4em\relax Clarendon Press, 1953.

\bibitem{young1938}
G.~Young and A.~Householder, ``Discussion of a Set of Points in Terms of Their
  Mutual Distances,'' \emph{Psychometrika}, vol.~3, no.~1, pp. 19--22, 1938.

\bibitem{kruskal1964}
J.~B. Kruskal, ``Multidimensional Scaling by Optimizing Goodness of Fit to a
  Nonmetric Hypothesis,'' \emph{Psychometrika}, vol.~29, no.~1, pp. 1--27,
  1964.

\bibitem{gower1982}
J.~C. Gower, ``Euclidean Distance Geometry,'' \emph{Math. Sci.}, vol.~7, pp. 1--14, 1982.

\bibitem{gower1}
------, ``Properties of {E}uclidean and non-{E}uclidean Distance Matrices,''
  \emph{Linear Algebra Appl.}, vol.~67, pp. 81--97, 1985.

\bibitem{Glunt1990}
W.~Glunt, T.~L. Hayden, S.~Hong, and J.~Wells, ``An Alternating Projection
  Algorithm for Computing the Nearest {E}uclidean Distance Matrix,''
  \emph{SIAM J. Matrix Anal. Appl.}, vol.~11, no.~4, pp. 589--600,
  1990.

\bibitem{Hayden1990}
T.~L. Hayden, J.~Wells, W.-M. Liu, and P.~Tarazaga, ``The Cone of Distance
  Matrices,'' \emph{Linear Algebra Appl.}, vol. 144, no.~0, pp.
  153--169, 1990.

\bibitem{Dattorro:2011wa}
J.~Dattorro, \emph{{Convex Optimization {\&} Euclidean Distance
  Geometry}}.\hskip 1em plus 0.5em minus 0.4em\relax Meboo, 2011.

\bibitem{trosset1998}
M.~W. Trosset, ``Applications of Multidimensional Scaling to Molecular
  Conformation,'' \emph{Comp. Sci. Stat.}, vol.~29, pp.
  148--152, 1998.

% \bibitem{Liberti:2011fs}
% L.~Liberti, C.~Lavor, A.~Mucherino, and N.~Maculan, ``{Molecular Distance
%   Geometry Methods: from Continuous to Discrete},'' \emph{Int. T. Oper. Res.}, vol.~18, no.~1, pp. 33--51, Jan. 2011.

\bibitem{Holm:1993dx}
L.~Holm and C.~Sander, ``{Protein Structure Comparison by Alignment of Distance
  Matrices},'' \emph{J. Mol. Biol.}, vol. 233, no.~1, pp.
  123--138, Sep. 1993.

\bibitem{Tenenbaum2000}
J.~B. Tenenbaum, V.~De Silva, and J.~C. Langford, ``A Global Geometric
  Framework for Nonlinear Dimensionality Reduction,'' \emph{Science}, vol. 290,
  no. 5500, pp. 2319--2323, 2000.

\bibitem{jain2004}
V.~Jain and L.~Saul, ``Exploratory Analysis and Visualization of Speech and
  Music by Locally Linear Embedding,'' in \emph{IEEE Trans. Acoust., Speech, Signal
  Process.}, vol.~3, 2004.

\bibitem{Demaine:2009dw}
E.~D. Demaine, F.~Gomez-Martin, H.~Meijer, D.~Rappaport, P.~Taslakian, G.~T.
  Toussaint, T.~Winograd, and D.~R. Wood, ``{The Distance Geometry of Music},''
  \emph{Comput. Geom.}, vol.~42, no.~5, pp. 429--454, Jul. 2009.

\bibitem{So:2007cz}
A.~M.-C. So and Y.~Ye, ``{Theory of Semidefinite Programming for Sensor Network
  Localization},'' \emph{Math. Program.}, vol. 109, no. 2-3, pp. 367--384, Mar.
  2007.

\bibitem{Crocco:2012eu}
M.~Crocco, A.~D. Bue, and V.~Murino, ``{A Bilinear Approach to the Position
  Self-Calibration of Multiple Sensors},'' \emph{IEEE Trans. Signal Process.},
  vol.~60, no.~2, pp. 660--673, 2012.

\bibitem{Pollefeys:2008ho}
M.~Pollefeys and D.~Nister, ``{Direct Computation of Sound and Microphone
  Locations from Time-Difference-of-Arrival Data},'' in \emph{Proc. Intl.
  Workshop on HSC}.\hskip 1em plus 0.5em minus 0.4em\relax Las Vegas,
  2008, pp. 2445--2448.

% \bibitem{Gaubitch:2013km}
% N.~D. Gaubitch, W.~B. Kleijn, and R.~Heusdens, ``{Auto-localization in Ad-hoc
%   Microphone Arrays},'' in \emph{Proc. IEEE ICASSP}\hskip 1em plus 0.5em minus 0.4em\relax Vancouver,
%   2013, pp. 106--110.

% \bibitem{Raykar:2004is}
% V.~C. Raykar and R.~Duraiswami, ``{Automatic Position Calibration of Multiple
%   Microphones},'' in \emph{Proc. IEEE ICASSP}\hskip 1em plus 0.5em minus 0.4em\relax Montreal, 2004, pp.
%   69--72.

\bibitem{Dokmanic:2014tc}
I.~Dokmani{\'c}, L.~Daudet, and M.~Vetterli, ``{How to Localize Ten Microphones
  in One Fingersnap},'' \emph{Proc. EUSIPCO}, 2014.

% \bibitem{Thrun:2005xx}
% S.~Thrun, ``{Affine Structure from Sound},'' in \emph{Proc. NIPS}.\hskip 1em plus 0.5em minus 0.4em\relax Cambridge,
%   MA: MIT Press, 2005.

\bibitem{Schonemann:1970wd}
P.~H. Sch{\"o}nemann, ``{On Metric Multidimensional Unfolding},''
  \emph{Psychometrika}, vol.~35, no.~3, pp. 349--366, 1970.

\bibitem{Gujarathi:2014cz}
S.~R. Gujarathi, C.~L. Farrow, C.~Glosser, L.~Granlund, and P.~M. Duxbury,
  ``{Ab-Initio Reconstruction of Complex Euclidean Networks in Two
  Dimensions},'' \emph{Physical Review E}, vol.~89, no.~5, 2014.

\bibitem{Krislock:2012xx}
N.~Krislock and H.~Wolkowicz, ``{Euclidean Distance Matrices and
  Applications},'' in \emph{Handbook on Semidefinite, Conic and Polynomial
  Optimization}.\hskip 1em plus 0.5em minus 0.4em\relax Boston, MA: Springer
  US, Jan. 2012, pp. 879--914.

\bibitem{Mucherino:2012hw}
A.~Mucherino, C.~Lavor, L.~Liberti, and N.~Maculan, \emph{{Distance Geometry:
Theory, Methods, and Applications}}. \hskip 1em plus 0.5em minus 0.4em\relax New York, NY:
  Springer Science {\&} Business Media, Dec. 2012.

\bibitem{Schonemann:1964tj}
P.~H. Sch{\"o}nemann, ``{A Solution of the Orthogonal Procrustes Problem With
  Applications to Orthogonal and Oblique Rotation},'' Ph.D. dissertation,
  University of Illinois at Urbana-Champaign, 1964.
  
\bibitem{Keshavan:2010bt}
R.~H. Keshavan, A.~Montanari, and S.~Oh, ``{Matrix Completion From a Few
  Entries},'' \emph{IEEE Trans. Inf. Theory}, vol.~56, no.~6, pp. 2980--2998,
  Jun. 2010.

\bibitem{Keshavan:2012tb}
------, ``{Matrix Completion from Noisy Entries},'' \emph{arXiv}, Apr. 2012.


\bibitem{dur07}
N.~Duric, P.~Littrup, L.~Poulo, A.~Babkin, R.~Pevzner, E.~Holsapple, O.~Rama,
  and C.~Glide, ``Detection of Breast Cancer with Ultrasound Tomography: First
  Results with the Computed Ultrasound Risk Evaluation (CURE) Prototype,''
  \emph{J. Med. Phys.}, vol.~34, no.~2, pp. 773--785, 2007.

\bibitem{parhizkar:2013a}
R.~Parhizkar, A.~Karbasi, S.~Oh, and M.~Vetterli, ``Calibration Using Matrix
  Completion with Application to Ultrasound Tomography,'' \emph{IEEE Trans.
  Signal Process.}, July 2013.

\bibitem{Borg2005}
I.~Borg and P.~Groenen, \emph{Modern Multidimensional Scaling: Theory and
  Applications}.\hskip 1em plus 0.5em minus 0.4em\relax Springer, 2005.

\bibitem{kruskal1964b}
J.~B. Kruskal, ``Nonmetric Multidimensional Scaling: A Numerical Method,''
  \emph{Psychometrika}, vol.~29, no.~2, pp. 115--129, 1964.

\bibitem{DeLeeuw1977}
J.~De~Leeuw, ``Applications of Convex Analysis to Multidimensional Scaling,''
  in \emph{Recent Developments in Statistics}, J.~Barra, F.~Brodeau, G.~Romier,
  and B.~V. Cutsem, Eds.\hskip 1em plus 0.5em minus 0.4em\relax North Holland
  Publishing Company, 1977, pp. 133--146.

\bibitem{mathar1991}
R.~Mathar and P.~J.~F. Groenen, ``Algorithms in Convex Analysis Applied to
  Multidimensional Scaling,'' in \emph{Symbolic-numeric data analysis and
  learning}, E.~Diday and Y.~Lechevallier, Eds.\hskip 1em plus 0.5em minus
  0.4em\relax Nova Science, 1991, pp. 45--56.

\bibitem{guttman1968}
L.~Guttman, ``A General Nonmetric Technique for Finding the Smallest Coordinate
  Space for a Configuration of Points,'' \emph{Psychometrika}, vol.~33, no.~4,
  pp. 469--506, 1968.

\bibitem{takane1977}
Y.~Takane, F.~Young, and J.~De~Leeuw, ``Nonmetric Individual Differences
  Multidimensional Scaling: An Alternating Least Squares Method with Optimal
  Scaling Features,'' \emph{Psychometrika}, vol.~42, no.~1, pp. 7--67, 1977.

\bibitem{gaffke1989}
N.~Gaffke and R.~Mathar, ``A Cyclic Projection Algorithm via Duality,''
  \emph{Metrika}, vol.~36, no.~1, pp. 29--54, 1989.

\bibitem{rezathesis13}
R.~Parhizkar, ``Euclidean Distance Matrices: Properties, Algorithms and
  Applications,'' Ph.D. dissertation, Ecole Polytechnique Federale de Lausanne
  (EPFL), 2013.

% \bibitem{malone2000}
% S.~W. Malone and M.~W. Trosset, ``A Study of the Stationary Configurations of
%   the s-stress Criterion for Metric Multidimensional Scaling,'' Duke University,
%   Tech. Rep., 2000.

\bibitem{Browne1987}
M.~Browne, ``The Young-Householder Algorithm and the Least Squares
  Multidimensional Scaling of Squared Distances,'' \emph{J. Classif.}, vol.~4,
  no.~2, pp. 175--190, 1987.

\bibitem{gluntEmbedding1991}
W.~Glunt, T.~L. Hayden, and W.-M. Liu, ``The Embedding Problem for Predistance
  Matrices,'' \emph{Bull. Math. Biol.}, vol.~53, no.~5, pp.
  769--796, 1991.


\bibitem{Biswas:2006cm}
P.~Biswas, T.~C. Liang, K.~C. Toh, Y.~Ye, and T.~C. Wang, ``{Semidefinite
  Programming Approaches for Sensor Network Localization With Noisy Distance
  Measurements},'' \emph{{IEEE} Trans. Autom. Sci. Eng.}, vol.~3, no.~4, pp. 360--371, 2006.

% \bibitem{Krislock:2010ga}
% N.~Krislock and H.~Wolkowicz, ``{Explicit Sensor Network Localization using
%   Semidefinite Representations and Facial Reductions},'' \emph{SIAM J.
%   Optimiz.}, vol.~20, no.~5, pp. 2679--2708, 2010.

\bibitem{cvx}
M.~Grant and S.~Boyd, ``{CVX}: Matlab Software for Disciplined Convex
  Programming, version 2.1,'' \url{http://cvxr.com/cvx}, Mar. 2014.

\bibitem{gb08}
------, ``Graph Implementations for Nonsmooth Convex Programs,'' in
  \emph{Recent Advances in Learning and Control}, ser. Lecture Notes in Control
  and Information Sciences, V.~Blondel, S.~Boyd, and H.~Kimura, Eds.\hskip 1em
  plus 0.5em minus 0.4em\relax Springer-Verlag Limited, 2008, pp. 95--110,
  \url{http://stanford.edu/~boyd/graph_dcp.html}.

\bibitem{Boutin:2004gb}
M.~Boutin and G.~Kemper, ``{On Reconstructing N-Point Configurations from the
  Distribution of Distances or Areas},'' \emph{Adv. Appl. Math.}, vol.~32,
  no.~4, pp. 709--735, May 2004.

\bibitem{Allen:1979ua}
J.~B. Allen and D.~A. Berkley, ``{Image Method for Efficiently Simulating
  Small-room Acoustics},'' \emph{J. Acoust. Soc. Am.}, vol.~65, no.~4, pp.
  943--950, 1979.

% \bibitem{Borish:1984uu}
% J.~Borish, ``{Extension of the Image Model to Arbitrary Polyhedra},'' \emph{J.
%   Acoust. Soc. Am.}, vol.~75, no.~6, pp. 1827--1836, 1984.

\bibitem{Millane:1990pt}
R.~P. Millane, ``{Phase Retrieval in Crystallography and Optics},'' \emph{J.
  Opt. Soc. Am. A}, vol.~7, no.~3, pp. 394--411, Mar. 1990.

\bibitem{Beavers:1989tj}
W.~Beavers, D.~E. Dudgeon, J.~W. Beletic, and M.~T. Lane, ``{Speckle Imaging
  Through the Atmosphere},'' \emph{Linconln Lab. J.}, vol.~2,
  pp. 207--228, 1989.

% \bibitem{Jaganathan:2012ta}
% K.~Jaganathan, S.~Oymak, and B.~Hassibi, ``{Recovery of Sparse 1-D Signals from
%   the Magnitudes of their Fourier Transform},'' \emph{IEEE ISIT}, pp. 1473--1477,
%   2012.

\bibitem{Lemke:2002um}
S.~S. Skiena, W.~D. Smith, and P.~Lemke, ``{Reconstructing Sets from Interpoint
  Distances},'' in \emph{ACM SCG}.\hskip 1em plus
  0.5em minus 0.4em\relax 1990, pp. 332--339.

% \bibitem{nash56}
% J.~Nash, ``The Embedding Problem for {R}iemannian Manifolds,'' \emph{Ann.
% Math.}, vol.~63, pp. 20--63, 1956.

\bibitem{Parhizkar:2014kn}
R.~Parhizkar, I.~Dokmani{\'c}, and M.~Vetterli, ``{Single-Channel Indoor Microphone Localization},'' \emph{Proc. IEEE ICASSP}, Florence, 2014.


\end{thebibliography}
% \linespread{1.41}

\end{document}